\documentclass[%
reprint,
superscriptaddress,
amsmath,amssymb,
pof,
]{revtex4-2}

\usepackage{graphicx}
\usepackage{dcolumn}
\usepackage{bm}
\usepackage[hidelinks]{hyperref}
\usepackage{cleveref}
\usepackage{tikz}
\usetikzlibrary{tikzmark,decorations.pathreplacing,calligraphy, arrows.meta, positioning, fit, calc}
\usepackage{mathtools}
\usepackage[dvipsnames]{xcolor}

\definecolor{forestgreen}{RGB}{34,139,34}
\definecolor{burntorange}{RGB}{204, 85, 0}

\definecolor{cinnamon}{rgb}{0.82, 0.41, 0.12}
\definecolor{electricultramarine}{rgb}{0.25, 0.0, 1.0}
\definecolor{green}{rgb}{0, 0.5, 0}
\definecolor{mplpurple}{rgb}{0.501960784314,0.0,0.501960784314}

\begin{document}

\title[Generalized OReg LBM]{Generalized Onsager-Regularized Lattice Boltzmann framework}

\author{Anirudh Jonnalagadda}
\email[Corresponding author: ]{anirudh.jonnalagadda@iit.it}
\affiliation{Computational mOdelling of NanosCalE and bioPhysical sysTems (CONCEPT) Lab, Istituto Italiano di Tecnologia, 16152 Genova, Italy}
\affiliation{Center for Life Nano- \& Neuro-Science, Fondazione Istituto Italiano di Tecnologia, Viale Regina Elena 295, Rome, 00161, Italy}
\author{Walter Rocchia}
\affiliation{Computational mOdelling of NanosCalE and bioPhysical sysTems (CONCEPT) Lab, Istituto Italiano di Tecnologia, 16152 Genova, Italy}
\author{Sauro Succi}
\affiliation{Center for Life Nano- \& Neuro-Science, Fondazione Istituto Italiano di Tecnologia, Viale Regina Elena 295, Rome, 00161, Italy}
\affiliation{Istituto per le Applicazioni del Calcolo, Consiglio Nazionale delle Ricerche, Via dei Taurini 19, Rome, 00185, Italy}
\affiliation{Department of Physics, Harvard University, 17 Oxford St., Cambridge, 02138, MA, USA}

\begin{abstract}
\noindent
This work presents a novel strategy to address Navier-Stokes modelling errors arising on first-nearest neighbour lattice Boltzmann (LB) methods and introduces fully local corrections through Onsager-Regularized (OReg) non-equilibrium populations. The proposed mechanism, which admits partially and completely corrected OReg models, is used to develop representative partially and completely corrected models for the six-moment-constrained guided equilibrium (GEq) representation on the D2Q9 lattice. The former realization only addresses compatibility condition violations and improves the accuracy by two/four orders of magnitude at reference/arbitrary lattice temperatures respectively, while the latter additionally corrects stress tensor modelling errors, resulting in a fully corrected exact model. Numerical benchmarks of the corrected schemes demonstrate improved accuracy and stability in comparison to the Lattice-BGK, regularized, and uncorrected OReg-GEq schemes thus presenting a promising avenue for OReg based thermohydrodynamic extensions.
\end{abstract}


\maketitle
\section{Introduction}
\label{sec:intro}
\noindent
The lattice Boltzmann (LB) method is a mesoscopic computational framework in which molecular collisions emulated at individual grid nodes are followed by the propagation of information to corresponding nearest-neighbours \citep{succi2018lattice, kruger}.
For single relaxation time (SRT) schemes, these operations are mathematically expressed through the lattice Boltzmann equation (LBE):
\vspace{-0.5em}
\begin{subequations}
    \label{eq:lbe}
	\begin{align}
		f_i^*(x_\alpha, t) = f_i^{(eq)}(x_\alpha, t) &+ \left(1-\frac{1}{\tau}\right)f_i^{(neq)}(x_\alpha, t),\\
		f_i^*(x_\alpha, t) = f_i&(x_\alpha + c_{i_\alpha}\Delta t, t + \Delta t)
	\end{align}
\end{subequations}
where $f_i^*$ represent the post-collision populations and the population set $\lbrace f_i : \, i \in [0, Q) \rbrace $ is defined on stencils comprising of $Q$ lattice entries each having a $D$-dimensional lattice velocity $c_{i_\alpha}$.
The quantities $f_i^{(eq)}$ and $f_i^{(neq)} = \left(f_i - f_i^{(eq)}\right)$ correspond to the equilibrium and non-equilibrium populations respectively, 	while the BGK relaxation time, $\tau = \nu/\theta$, is the ratio of the lattice viscosity, $\nu$, to the reduced temperature $\theta$.
The corresponding macroscopic mass and momentum densities are obtained from the lattice populations as $\displaystyle\rho = \sum\limits_{i} f_i$ and $\displaystyle\rho u_\alpha = \sum\limits_{i} c_{i_\alpha}f_i$ respectively.
The LB algorithm presented in \Cref{eq:lbe}, which inherently involves localized non-linear collisions and linear non-local propagations, is particularly well suited to harness the power of modern compute infrastructures \citep{exascaleLB}.
Indeed, successful applications in numerous scientific and commercial settings have led to a significant uptake of LB schemes as an alternative computational fluid dynamics methodology.

\begin{figure}[t]
    {
        \begin{tikzpicture}[>=latex]
        \def\A{(-1,-1)}; \def\B{(1,-1)};
        \def\C{(1,1)}; \def\D{(-1,1)};

        \def\E{(0,-1)}; \def\F{(0,1)};
        \def\G{(-1,0)}; \def\H{(1,0)};

        \draw[gray] \A--\B--\C--\D--cycle; 
        \draw[gray] \E--\F; 
        \draw[gray] \G--\H; 

        \draw[->, line width=0.35mm] (0,0) -- ( -1, 0) node[left ] { \tiny 3 }; 
        \draw[->, line width=0.35mm] (0,0) -- (  1, 0) node[right] { \tiny 1 };  
        \draw[->, line width=0.35mm] (0,0) -- (  0,-1) node[below] { \tiny 4 }; 
        \draw[->, line width=0.35mm] (0,0) -- (  0, 1) node[above] { \tiny 2 };  
        \draw[->, line width=0.35mm] (0,0) -- ( -1,-1) node[below left ] {\tiny 7 }; 
        \draw[->, line width=0.35mm] (0,0) -- (  1,-1) node[below right] {\tiny 8 };  
        \draw[->, line width=0.35mm] (0,0) -- (  1, 1) node[above right] {\tiny 5 };  
        \draw[->, line width=0.35mm] (0,0) -- ( -1, 1) node[above left ] {\tiny 6 };  

        \fill[white] (0,0) circle (3pt);
        \draw (0, 0) node {\tiny 0};
        \end{tikzpicture}
        \begin{picture}(0,0)
            \put(-67.5, -10){(a) D2Q9}
        \end{picture}
    }
    {
        \begin{tikzpicture}[>=latex]
            \def\A{(-1,-1,-1)}; \def\P{(-1, -1, 0)}; \def\E{(-1,-1, 1)};
            \def\B{( 1,-1,-1)}; \def\Q{( 1, -1, 0)}; \def\F{( 1,-1, 1)};
            \def\C{( 1, 1,-1)}; \def\R{( 1,  1, 0)}; \def\G{( 1, 1, 1)};
            \def\D{(-1, 1,-1)}; \def\S{(-1,  1, 0)}; \def\H{(-1, 1, 1)};
            \def\L{(-1, 0, 1)}; \def\M{(-1, 0,-1)}; \def\N{(1, 0,-1)}; \def\O{( 1, 0, 1)};

            \def\AA{(0,-1, 1)}; \def\BB{(0,-1,-1)}; \def\CC{(0, 1,-1)}; \def\DD{(0, 1, 1)};

            \draw[gray] \A--\B--\C--\D--cycle; 
            \draw[gray] \E--\F--\G--\H--cycle; 
            \draw[gray] \P--\Q--\R--\S--cycle; 
            \draw[gray] \A--\E; 
            \draw[gray] \B--\F; 
            \draw[gray] \C--\G; 
            \draw[gray] \D--\H; 
            \draw[gray] \L--\M--\N--\O--cycle; 
            \draw[gray] \AA--\BB--\CC--\DD--cycle; 

            \draw[->, thick, green] (0,0,0) -- ( 1, 0, 0) node[right, xshift=-2pt, yshift=-1pt] { \tiny 1 }; 
            \draw[->, thick, green] (0,0,0) -- (-1, 0, 0) node[left , xshift= 1.5pt] { \tiny 2 }; 
            \draw[->, thick, green] (0,0,0) -- ( 0, 1, 0) node[above, xshift=-1pt, yshift=-2pt] { \tiny 3 }; 
            \draw[->, thick, green] (0,0,0) -- ( 0,-1, 0) node[below, xshift=1pt, yshift=2pt] { \tiny 4 }; 
            \draw[->, thick, green] (0,0,0) -- ( 0, 0, 1) node[above, xshift=-2.5pt, yshift=-3pt] { \tiny 5 }; 
            \draw[->, thick, green] (0,0,0) -- ( 0, 0,-1) node[above, xshift=-2.5pt, yshift=-3pt] { \tiny 6 }; 
            \draw[->, thick, cinnamon, dotted, line width = 0.35mm] (0,0,0) -- ( 1, 0,-1) node[right] { \tiny 15 };  
            \draw[->, thick, cinnamon, dotted, line width = 0.35mm] (0,0,0) -- (-1, 0, 1) node[left ] { \tiny 16 };  
            \draw[->, thick, cinnamon, dotted, line width = 0.35mm] (0,0,0) -- ( 1, 0, 1) node[below, xshift= 3pt, yshift= 3pt,] { \tiny 9  };  
            \draw[->, thick, cinnamon, dotted, line width = 0.35mm] (0,0,0) -- (-1, 0,-1) node[above, xshift=-4pt, yshift=-3pt,] { \tiny 10 };  
            \draw[->, thick, cinnamon, dotted, line width = 0.35mm] (0,0,0) -- (-1,-1, 0) node[left , xshift= 2pt, yshift= 2pt,] { \tiny 8 };  
            \draw[->, thick, cinnamon, dotted, line width = 0.35mm] (0,0,0) -- ( 1, 1, 0) node[right, xshift=-3pt, yshift=-3pt,] { \tiny 7 };  
            \draw[->, thick, cinnamon, dotted, line width = 0.35mm] (0,0,0) -- ( 0, 1,-1) node[above,                          ] { \tiny 17 }; 
            \draw[->, thick, cinnamon, dotted, line width = 0.35mm] (0,0,0) -- ( 0,-1, 1) node[below,                          ] { \tiny 18 }; 
            \draw[->, thick, cinnamon, dotted, line width = 0.35mm] (0,0,0) -- (-1, 1, 0) node[left , xshift= 3pt, yshift= 3pt,] { \tiny 14 };  
            \draw[->, thick, cinnamon, dotted, line width = 0.35mm] (0,0,0) -- ( 1,-1, 0) node[right, xshift=-3pt, yshift=-3pt,] { \tiny 13 };  
            \draw[->, thick, cinnamon, dotted, line width = 0.35mm] (0,0,0) -- ( 0, 1, 1) node[above, xshift=-2.5pt, yshift=-3pt] { \tiny 11 };  
            \draw[->, thick, cinnamon, dotted, line width = 0.35mm] (0,0,0) -- ( 0,-1,-1) node[below, xshift= 2.5pt, yshift=3pt] { \tiny 12 };  
            \draw[->, line width = 0.35mm, blue] (0, 0, 0) -- ( 1, 1, 1) node[right, xshift=-3pt, yshift=-3pt] { \tiny 19 }; 
            \draw[->, line width = 0.35mm, blue] (0, 0, 0) -- (-1,-1,-1) node[left , xshift= 3pt, yshift= 3pt] { \tiny 20 }; 
            \draw[->, line width = 0.35mm, blue] (0, 0, 0) -- ( 1, 1,-1) node[above] { \tiny 21 }; 
            \draw[->, line width = 0.35mm, blue] (0, 0, 0) -- (-1,-1, 1) node[below] { \tiny 22 }; 
            \draw[->, line width = 0.35mm, blue] (0, 0, 0) -- ( 1,-1, 1) node[below] { \tiny 23 }; 
            \draw[->, line width = 0.35mm, blue] (0, 0, 0) -- (-1, 1,-1) node[above] { \tiny 24 }; 
            \draw[->, line width = 0.35mm, blue] (0, 0, 0) -- (-1, 1, 1) node[left ] { \tiny 25 }; 
            \draw[->, line width = 0.35mm, blue] (0, 0, 0) -- ( 1,-1,-1) node[right] { \tiny 26 }; 

            \fill[white] (0,0) circle (3pt);
            \draw (0, 0) node {\tiny 0};
        \end{tikzpicture}
        \begin{picture}(0,0)
        \put(-90, -10){(b) D3Q27}
        \end{picture}
    }

    \vspace{1.25em}

    {
        \begin{tikzpicture}[>=latex]
            \def\A{(-1,-1,-1)}; \def\P{(-1, -1, 0)}; \def\E{(-1,-1, 1)};
            \def\B{( 1,-1,-1)}; \def\Q{( 1, -1, 0)}; \def\F{( 1,-1, 1)};
            \def\C{( 1, 1,-1)}; \def\R{( 1,  1, 0)}; \def\G{( 1, 1, 1)};
            \def\D{(-1, 1,-1)}; \def\S{(-1,  1, 0)}; \def\H{(-1, 1, 1)};
            \def\L{(-1, 0, 1)}; \def\M{(-1, 0,-1)}; \def\N{(1, 0,-1)}; \def\O{( 1, 0, 1)};

            \def\AA{(0,-1, 1)}; \def\BB{(0,-1,-1)}; \def\CC{(0, 1,-1)}; \def\DD{(0, 1, 1)};

            \draw[gray] \A--\B--\C--\D--cycle; 
            \draw[gray] \E--\F--\G--\H--cycle; 
            \draw[gray] \P--\Q--\R--\S--cycle; 
            \draw[gray] \A--\E; 
            \draw[gray] \B--\F; 
            \draw[gray] \C--\G; 
            \draw[gray] \D--\H; 
            \draw[gray] \L--\M--\N--\O--cycle; 
            \draw[gray] \AA--\BB--\CC--\DD--cycle; 

            \draw[->, thick, green] (0,0,0) -- ( 1, 0, 0) node[right, xshift=-2pt, yshift=-1pt] { \tiny 1 }; 
            \draw[->, thick, green] (0,0,0) -- (-1, 0, 0) node[left , xshift= 1.5pt] { \tiny 2 }; 
            \draw[->, thick, green] (0,0,0) -- ( 0, 1, 0) node[above, xshift=-1pt, yshift=-2pt] { \tiny 3 }; 
            \draw[->, thick, green] (0,0,0) -- ( 0,-1, 0) node[below, xshift=1pt, yshift=2pt] { \tiny 4 }; 
            \draw[->, thick, green] (0,0,0) -- ( 0, 0, 1) node[above, xshift=-2.5pt, yshift=-3pt] { \tiny 5 }; 
            \draw[->, thick, green] (0,0,0) -- ( 0, 0,-1) node[above, xshift=-2.5pt, yshift=-3pt] { \tiny 6 }; 
            \draw[->, thick, cinnamon, dotted, line width = 0.35mm] (0,0,0) -- ( 1, 0,-1) node[right] { \tiny 15 };  
            \draw[->, thick, cinnamon, dotted, line width = 0.35mm] (0,0,0) -- (-1, 0, 1) node[left ] { \tiny 16 };  
            \draw[->, thick, cinnamon, dotted, line width = 0.35mm] (0,0,0) -- ( 1, 0, 1) node[below, xshift= 3pt, yshift= 3pt,] { \tiny 9  };  
            \draw[->, thick, cinnamon, dotted, line width = 0.35mm] (0,0,0) -- (-1, 0,-1) node[above, xshift=-4pt, yshift=-3pt,] { \tiny 10 };  
            \draw[->, thick, cinnamon, dotted, line width = 0.35mm] (0,0,0) -- (-1,-1, 0) node[left , xshift= 2pt, yshift= 2pt,] { \tiny 8 };  
            \draw[->, thick, cinnamon, dotted, line width = 0.35mm] (0,0,0) -- ( 1, 1, 0) node[right, xshift=-3pt, yshift=-3pt,] { \tiny 7 };  
            \draw[->, thick, cinnamon, dotted, line width = 0.35mm] (0,0,0) -- ( 0, 1,-1) node[above,                          ] { \tiny 17 }; 
            \draw[->, thick, cinnamon, dotted, line width = 0.35mm] (0,0,0) -- ( 0,-1, 1) node[below,                          ] { \tiny 18 }; 
            \draw[->, thick, cinnamon, dotted, line width = 0.35mm] (0,0,0) -- (-1, 1, 0) node[left , xshift= 3pt, yshift= 3pt,] { \tiny 14 };  
            \draw[->, thick, cinnamon, dotted, line width = 0.35mm] (0,0,0) -- ( 1,-1, 0) node[right, xshift=-3pt, yshift=-3pt,] { \tiny 13 };  
            \draw[->, thick, cinnamon, dotted, line width = 0.35mm] (0,0,0) -- ( 0, 1, 1) node[above, xshift=-2.5pt, yshift=-3pt] { \tiny 11 };  
            \draw[->, thick, cinnamon, dotted, line width = 0.35mm] (0,0,0) -- ( 0,-1,-1) node[below, xshift= 2.5pt, yshift=3pt] { \tiny 12 };  

            \fill[white] (0,0) circle (3pt);
            \draw (0, 0) node {\tiny 0};
        \end{tikzpicture}
        \begin{picture}(0,0)
        \put(-90, -10){(c) D3Q19}
        \end{picture}
    }
    {
        \begin{tikzpicture}[>=latex]
            \def\A{(-1,-1,-1)}; \def\P{(-1, -1, 0)}; \def\E{(-1,-1, 1)};
            \def\B{( 1,-1,-1)}; \def\Q{( 1, -1, 0)}; \def\F{( 1,-1, 1)};
            \def\C{( 1, 1,-1)}; \def\R{( 1,  1, 0)}; \def\G{( 1, 1, 1)};
            \def\D{(-1, 1,-1)}; \def\S{(-1,  1, 0)}; \def\H{(-1, 1, 1)};
            \def\L{(-1, 0, 1)}; \def\M{(-1, 0,-1)}; \def\N{(1, 0,-1)}; \def\O{( 1, 0, 1)};

            \def\AA{(0,-1, 1)}; \def\BB{(0,-1,-1)}; \def\CC{(0, 1,-1)}; \def\DD{(0, 1, 1)};

            \draw[gray] \A--\B--\C--\D--cycle; 
            \draw[gray] \E--\F--\G--\H--cycle; 
            \draw[gray] \P--\Q--\R--\S--cycle; 
            \draw[gray] \A--\E; 
            \draw[gray] \B--\F; 
            \draw[gray] \C--\G; 
            \draw[gray] \D--\H; 
            \draw[gray] \L--\M--\N--\O--cycle; 
            \draw[gray] \AA--\BB--\CC--\DD--cycle; 

            \draw[->, thick, green] (0,0,0) -- ( 1, 0, 0) node[right, xshift=-2pt, yshift=-1pt] { \tiny 1 }; 
            \draw[->, thick, green] (0,0,0) -- (-1, 0, 0) node[left , xshift= 1.5pt] { \tiny 2 }; 
            \draw[->, thick, green] (0,0,0) -- ( 0, 1, 0) node[above, xshift=-1pt, yshift=-2pt] { \tiny 3 }; 
            \draw[->, thick, green] (0,0,0) -- ( 0,-1, 0) node[below, xshift=1pt, yshift=2pt] { \tiny 4 }; 
            \draw[->, thick, green] (0,0,0) -- ( 0, 0, 1) node[above, xshift=-2.5pt, yshift=-3pt] { \tiny 5 }; 
            \draw[->, thick, green] (0,0,0) -- ( 0, 0,-1) node[above, xshift=-2.5pt, yshift=-3pt] { \tiny 6 }; 
            \draw[->, thick, cinnamon, dotted, line width = 0.35mm] (0, 0, 0) -- ( 1, 1, 1) node[right, xshift=-3pt, yshift=-3pt] { \tiny 7 }; 
            \draw[->, thick, cinnamon, dotted, line width = 0.35mm] (0, 0, 0) -- (-1,-1,-1) node[left , xshift= 3pt, yshift= 3pt] { \tiny 8 }; 
            \draw[->, thick, cinnamon, dotted, line width = 0.35mm] (0, 0, 0) -- ( 1, 1,-1) node[above] { \tiny 9 }; 
            \draw[->, thick, cinnamon, dotted, line width = 0.35mm] (0, 0, 0) -- (-1,-1, 1) node[below] { \tiny 10 }; 
            \draw[->, thick, cinnamon, dotted, line width = 0.35mm] (0, 0, 0) -- ( 1,-1, 1) node[below] { \tiny 11 }; 
            \draw[->, thick, cinnamon, dotted, line width = 0.35mm] (0, 0, 0) -- (-1, 1,-1) node[above] { \tiny 12 }; 
            \draw[->, thick, cinnamon, dotted, line width = 0.35mm] (0, 0, 0) -- (-1, 1, 1) node[left ] { \tiny 13 }; 
            \draw[->, thick, cinnamon, dotted, line width = 0.35mm] (0, 0, 0) -- ( 1,-1,-1) node[right] { \tiny 14 }; 

            \fill[white] (0,0) circle (3pt);
            \draw (0, 0) node {\tiny 0};

        \end{tikzpicture}
        \begin{picture}(0,0)
        \put(-90, -10){(d) D3Q15}
        \end{picture}
    }
    \vspace{1em}
    \caption{\label{fig:lattice-representation} 2D and 3D Standard lattices (color online).}
    \vspace{-1em}
\end{figure}
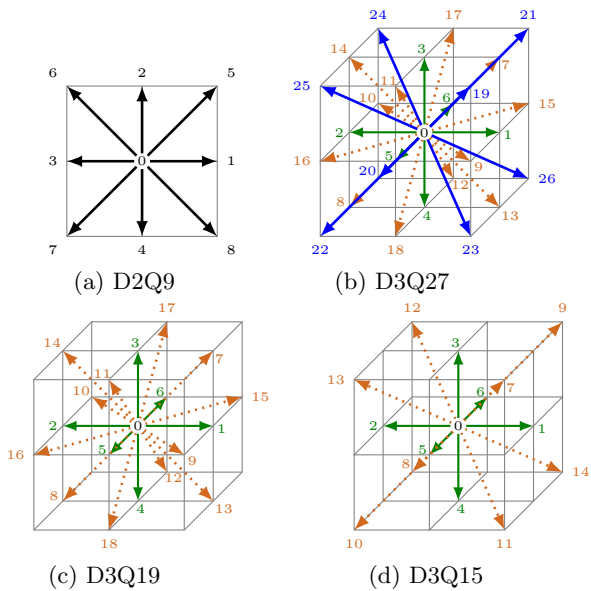

Traditionally, LB schemes have operated with standard, first-nearest-neighbour discrete velocity spaces (\Cref{fig:lattice-representation}).
However, since such lattices employ only three discrete speeds, $c_{i_\alpha} \in$ (0, $\pm$ 1), certain higher-order moments degenerate into lower-order representations.
As a result, standard-lattice LB models lose Galilean invariance, and can, therefore, only approximate the Navier-Stokes (NS) equations \citep{qian1992, qian1993, karlinasinari}.
Consequently, such models are best suited for isothermal subsonic applications operating at the standard lattice reference temperature $\theta_0 = 1/3$.
Indeed, for flows characterized by appreciable speeds and/or thermal fluctuations, the NS modelling errors arising on standard lattices produce grossly inaccurate results and often yield catastrophically unstable simulations.

Two strategies have been widely explored for enabling simulations of high speed flow phenomena within the LB framework.
The first identifies the under-represented standard lattice velocity space as the source of the NS modelling errors, and takes the intuitive route of employing larger, multispeed, velocity sets \citep{alexanderchensterling,kataokatsutahara,lihewangtao, watari, karlinlatticesPRE, karlinlatticesPRL, shanlattices2016, atifThermalLBM, praveenThermalLBM, atif2025Transonic}.
However, multispeed lattices are accurate only over small ranges of temperature \citep{frapollithesis} and, due to the large number of discrete velocities required to completely recover compressible NS dynamics, incur significantly high computational costs.
In the second, more direct, strategy, NS modelling errors are evaluated using the Chapman-Enskog (CE) multi-scale expansion procedure, and are eliminated by injecting correction populations into the LBE either via modified equilibrium representations \citep{SaadatDorschnerKarlin2021, SaadatAliDorschnerKarlin2021} or as explicit source terms \citep{PrasianakisKarlin2007, PrasianakisKarlin2008, PrasianakisKarlinMantzarasBoulouchosPRE2009, thermalFlowsStdLattices2012, fengSagautTao2015, SaadatBoschKarlinPRE2019, sagaut2020D3Q19r,fengHybridRecreg2019, hosseini2020CorrectionReview}.
However, it is important to note that the complexity of the correction populations are directly influenced by the analytical representation of the NS modelling errors.
Specifically, using closure procedures based on the evolution equations for the stress tensor and heat flux vector in the CE expansion, these errors manifest in terms of the gradients of higher-order equilibrium moments.
Usually, these error contributions, and the corresponding correction populations, can only be evaluated non-locally through strided memory accesses.
Thus the direct approach erodes the LB ethos of locality of computations and compromises the computational efficiency of standard lattices. Furthermore, such non-local corrections are intrinsically incompatible with advanced LB algorithms that aim to efficiently utilize modern hardware \citep{AAstreaming, esoterictwist}.

It can be appreciated that an ideal compressible LB model would endeavour not only to retain the computational advantages of standard lattices, but also to allow for fully local representations of the correction terms.
Here, the first effort towards this ideal is presented in the context of two-dimensional isothermal flow scenarios via the Onsager-Regularized (OReg) LB scheme \citep{jonnalagaddaJHT2021, jonnalagaddaPRE2021, jonnalagadda2023NHTB, davideThesis, jonnalagaddaPRR2025}.

The proposed correction mechanism differs from existing approaches in the following ways.
Firstly, correction populations are obtained through the OReg description of the non-equilibrium populations.
To the best of the authors' knowledge, this is the first time corrections to LB modelling errors explicitly employ non-equilibrium descriptions.
The next distinction lies in the significantly simpler mathematical representation of the OReg modelling errors.
Specifically, since the momentum transport equation in the CE expansion is closed by directly evaluating the stress tensor in terms of the OReg non-equilibrium populations, the corresponding OReg modelling errors have the generic product form $\displaystyle\widetilde{\mathrm{E}} = \mathrm{X}_{\alpha\beta}\cdot\phi(\Phi_{\alpha\beta}')$, where, $\mathrm{X}_{\alpha\beta} = \mathrm{S}_{\alpha\beta}/\theta$, with $\mathrm{S}_{\alpha\beta} = (\partial_\alpha u_\beta + \partial_\beta u_\alpha)/2$ being the strain rate tensor, is the non-equilibrium thermodynamic force associated with viscous processes, and $\phi$ is a scalar/vector function of the deviations of equilibrium moments from their Maxwell-Boltzmann (MB) counterparts $\Phi_{\alpha\beta}'$.
In contrast, and as mentioned earlier, existing correction mechanisms serve to correct the modelling errors that take the form of gradients of higher-order equilibrium moments. It is noteworthy that $\Phi_{\alpha\beta}'$ is an explicit function of the macroscopic field variables, and can be trivially evaluated once the equilibrium representation is fixed.
Similarly, $\mathrm{X}_{\alpha\beta}$ can be locally evaluated through the definition of the incompressible stress tensor \citep{jonnalagaddaJHT2021,jonnalagaddaPRE2021,jonnalagadda2023NHTB}.
Thus, the third, and most important, distinction of the proposed OReg-based correction scheme is its amenability to yield fully-local correction populations.
The last differentiating aspect of the proposed correction mechanism lies in the type of modelling errors that are eliminated.
It is important to note that the isothermal NS modelling errors on standard lattices emerge from two mutually independent sources, namely, erroneous stress-tensor representations as well as compatibility condition violations \citep{jonnalagaddaPRR2025}.
The proposed OReg-based correction procedure can address either of the two error sources and is, thus, able to yield partially or completely corrected isothermal OReg kinetic models.
Note that, since OReg schemes already yield an order of accuracy improvement over traditional LB methods \citep{jonnalagaddaPRR2025}, the additional accuracy improvement with such partially corrected models are expected to be useful in scenarios involving small deviations from the lattice reference temperature.
Similarly, the fully corrected models serve as stepping stones towards OReg-based standard lattice compressible LB models.

The remainder of the manuscript is laid out as follows:
\Cref{sec:OReg} describes the uncorrected isothermal OReg scheme along with its hydrodynamic limit and associated NS modelling errors.
The OReg-correction framework is then presented in an equilibrium- and stencil-agnostic setting in \Cref{sec:genOReg}, and is subsequently deployed for the D2Q9 guided-equilibrium (GEq) representation \citep{PrasianakisKarlin2007, PrasianakisKarlin2008, PrasianakisKarlinMantzarasBoulouchosPRE2009} in \Cref{sec:guided-models} to develop a representative partially corrected model, that improves the accuracy of the OReg-GEq scheme from $\mathcal{O}(u)$ and $\mathcal{O}(u^3)$ at $\theta\neq 1/3$ and $\theta = 1/3$ to $\mathcal{O}(u^5)$ at all for isothermal lattice temperatures, along with a completely corrected, exact OReg-GEq model for NS macrodynamics.
In \Cref{sec:results}, the behaviour of the proposed corrected OReg-GEq models are evaluated against the uncorrected OReg-GEq scheme as well as the standard Lattice-BGK (LBGK) and regularized LB  schemes \citep{LATT2006} for two canonical quasi 1D problems and in a non-linear 2D setting, namely the rotated decaying shear wave, isothermal shocktube, and the doubly-periodic shear layer problems.
Lastly, \Cref{sec:conclusion} presents our conclusions.
\vspace{-1em}
\section{The Onsager Regularized Scheme}
\label{sec:OReg}
\noindent
The OReg scheme approaches LB through the lens of nonequilibrium thermodynamics and has empirically alluded to improved NS kinetic models on standard lattices \citep{jonnalagaddaPRE2021, davideThesis}.
Indeed, recent theoretical investigations have explained these empirical findings by demonstrating an order-of-magnitude accuracy improvement in isothermal dynamics, despite abandoning the customary assumption of compliance with compatibility conditions at the NS level \citep{jonnalagaddaPRR2025}.
In the isothermal OReg formulation for monatomic gases \cite{jonnalagaddaPRE2021,jonnalagaddaPRR2025}, the non-equilibrium populations used in \Cref{eq:lbe} are corrected from $f_i^{(neq)} = \left(f_i - f_i^{(eq)}\right)$ to $f_i^\text{OReg}$ and are given as:
\vspace{-1em}
\begin{multline}
    \label{eq:oreg}
    f_i^\text{OReg}
    =
    \frac{f_i^{(eq)}}{2\rho\theta^2}
    \Big(
        C_{i_\alpha}C_{i_\beta}
        -
	\frac{C_{i_\chi}^2}{D} \delta_{\alpha\beta}
    \Big)
    \\[-0.5em]
    \hspace{0.5em}
    \displaystyle
    \Bigg[
    \sum\limits_{k=0}^{Q-1}
    \Big(
        c_{k_\alpha}c_{k_\beta}
        -
        \frac{c_{k}^2}{D} \delta_{\alpha\beta}
    \Big)
	f_{k}^{(neq)}
    \Bigg],
\end{multline}
where $C_{i_\alpha} = \left(c_{i_\alpha} - u_\alpha\right)$ and $\delta_{\alpha\beta}$ is the Kronecker delta function.
With equilibrium representations that satisfy fundamental constraints of:
\begin{equation}
\label{eq:density-momentum}
 \displaystyle\sum\limits_{i}f_i^{(eq)} = \rho \text{ and }\displaystyle\sum\limits_{i} c_{i_\alpha}f_i^{(eq)} = \rho u_\alpha,
\end{equation}
and recover the MB equilibrium pressure tensor with an ideal gas equation of state,
\begin{equation}
	\label{eq:ideal-gas-eos-eq-pressure-tensor}
	\displaystyle \Pi_{\alpha\beta}^{(eq)} = \Pi_{\alpha\beta}^\text{MB} = \sum\limits_i c_{i_\alpha}c_{i_\beta} f_i^{(eq)} = \rho (u_\alpha u_\beta + \theta \delta_{\alpha\beta}),
\end{equation}
\noindent
a generic, assumption free, CE expansion of the OReg lattice Boltzmann equation at the NS level yields the following equations in the macroscopic limit \citep{jonnalagaddaPRR2025}:
\begin{subequations}
	\label{eq:CEmacroscopicEqs}
	\begin{align}
            \partial_t \rho
            +
            \partial_\alpha \left( \rho u_\alpha \right)
            &
            =
            \partial_\alpha
            \Bigg[
                \bigg(
                    1
                    -
                    \frac{\Delta t}{2\tau}
                \bigg)
                \psi_\alpha
            \Bigg],
            \label{eq:CEcontinuity}
            \\
            \partial_\alpha \left(\rho u_\alpha\right)
            +
            \partial_\beta \Pi_{\alpha\beta}^\text{MB}
            +
            \,
            \partial_\beta
            &
            \Bigg[
                \bigg( 1 - \frac{\Delta t}{2\tau} \bigg)
                \Pi_{\alpha\beta}
            \Bigg]
            \nonumber
            \\
            =
            -
            \frac{1}{\tau}\psi_{\alpha}
            &
            -
            \partial_t
            \Bigg[
                \bigg( 1 - \frac{\Delta t}{2\tau} \bigg)
                \psi_{\alpha}
            \Bigg]
            \label{eq:CEmomentum}.
        \end{align}
    \end{subequations}
The quantities $\psi_{\alpha}$ and $\Pi_{\alpha\beta}$ are the first order compatibility condition contributions and the OReg stress tensor representation respectively.
Explicitly, $\psi_{\alpha}$ and $\Pi_{\alpha\beta}$ are given as \citep{jonnalagaddaPRR2025}:
\begin{widetext}
	\vspace{-2em}
    \begin{subequations}
    	\label{eq:oreg-deviations}
        \begin{align}
            \psi_{\gamma}
            =
            \displaystyle
            \sum\limits_{i} c_{i_{\gamma}}f_i^\text{OReg}
            =
            -\tau
            \bigg(
                Q_{\alpha\beta\gamma}'
                -
                \frac{\delta_{\alpha\beta}}{D}
                    q_\gamma'
            \bigg)
            \mathrm{X}_{\alpha\beta}
            ,
            \phantom{~~~~~~~~~~~~~~~~~~~~}
            \label{eq:oreg-mom-compatibility}
            \\[-0.5em]
            \Pi_{\alpha\beta}
            =
            \displaystyle \sum\limits_{i} c_{i_{\alpha}}c_{i_{\beta}} f_i^\text{OReg}
            =
            \underbrace{
                -\tau\rho\theta
                \bigg(
                    \partial_\alpha u_\beta
                    +
                    \partial_\beta u_\alpha
                    -
                    \frac{2}{D}\partial_\chi u_\chi\delta_{\alpha\beta}
                \bigg)
             }_{\Pi_{\alpha\beta}^\text{NS}}
         \phantom{~~~~~~~~~~~~~~~~~~~~~~~~~~~~~~~~}
         \nonumber
         \\[-0.5em]
         \underbrace{
            -\tau
            \Big[
                R_{\alpha\beta\gamma\mu}'
                -
                u_\gamma Q_{\alpha\beta\mu}'
                -
                u_\mu Q_{\alpha\beta\gamma}'
                -
                \frac{\delta_{\gamma\mu}}{D}
                \big(
                    R_{\alpha\beta\chi\chi}'
                    -
                    2 u_\chi Q_{\alpha\beta\chi}'
                \big)
            \Big]
            \mathrm{X}_{\gamma\mu}
        }_{\widetilde{\Pi}_{\alpha\beta}}
        ,
        \label{eq:oreg-pressure-tensor}
        \end{align}
    \end{subequations}
    \vspace{-1em}
\end{widetext}

\noindent
where, $q_{\alpha}'$,  $Q_{\alpha\beta\gamma}'$, and $R_{\alpha\beta\gamma\mu}'$ represent, respectively, the deviations of the heat flux vector, the third-order and the fourth-order moments from their MB counterparts.
It can be appreciated that Equations (\ref{eq:CEmacroscopicEqs}) represent approximate mass and momentum conservation equations in which modelling errors arise due to non-zero values of the deviations $\left\lbrace \psi_\alpha, \widetilde{\Pi}_{\alpha\beta}\right\rbrace$.
In the next section, we describe a how these deviations are eliminated.
\section{Onsager-Regularized Correction Framework}
\label{sec:genOReg}
\noindent
To obtain the desired corrected NS representations, we begin by introducing correction populations, $\Psi_i$, in the OReg populations and define a generalized non-equilibrium representation as  $f_i^\text{GOReg} = f_i^\text{OReg} + \Psi_i$.
Note that such a generalization is equivalent to introducing a correcting source term $\mathrm{S}_\text{C} = \left(1-\frac{1}{\tau}\right) \Psi_i$ into \Cref{eq:lbe}.
Nevertheless, we retain the interpretation of augmented OReg non-equilibrium populations in the ensuing discussions.
To comply with the compatibility conditions and recover the NS stress tensor exactly, these newly defined $f_i^\text{GOReg}$ populations must satisfy the following conditions:
\vspace{-0.5em}
\begin{equation}
	\label{eq:genOReg-constraints}
	\displaystyle \sum\limits_i \big\lbrace 1, c_{i_{\alpha}}, c_{i_{\alpha}}c_{i_{\beta}}\big\rbrace f_i^\text{GOReg} = \left\lbrace 0, 0, \Pi_{\alpha\beta}^\text{NS} \right\rbrace,
	\vspace{-1em}
\end{equation}
\noindent
which directly impose the following constraints on $\Psi_i$:
\vspace{-0.75em}
\begin{equation}
	\label{eq:correction}
	\displaystyle
    \sum\limits_i \left\lbrace 1, c_{i_\alpha}, c_{i_\alpha}c_{i_\beta} \right\rbrace \Psi_i + \left\lbrace 0, \psi_\alpha, \widetilde{\Pi}_{\alpha\beta}\right\rbrace = \left\lbrace 0,0,0 \right\rbrace.
    \vspace{-1em}
\end{equation}
\noindent
It is noteworthy that \Cref{eq:correction}, which corresponds to 6 and 10 equations for the standard D2Q9 and D3Q15/19/27 lattices respectively, forms an under-determined system in both two- and three-dimensions.
Thus, an infinite family of corrected OReg models can be obtained.
Additionally, by relaxing the deviation set from $\left\lbrace \psi_\alpha, \widetilde{\Pi}_{\alpha\beta} \right\rbrace$ to $\left\lbrace \psi_\alpha, 0 \right\rbrace$, we can also obtain partially corrected approximate models having, in comparison to the uncorrected OReg scheme, further improved accuracies.
\vspace{1em}
\subsection{Corrected 2D Onsager Regularized models}
\label{sec:guided-models}
\noindent
As a concrete illustration of the proposed mechanism, we consider the OReg scheme in conjunction with the guided equilibrium obtained from an entropic minimization procedure that is constrained to exactly recover Equations (\ref{eq:density-momentum}) and (\ref{eq:ideal-gas-eos-eq-pressure-tensor}) on the D2Q9 lattice \citep{PrasianakisKarlin2007, PrasianakisKarlin2008, PrasianakisKarlinMantzarasBoulouchosPRE2009}.
Explicitly, the guided equilibrium is given as:
\begin{equation}
    \label{eq:guidedEq}
    f_i^{(eq)}= \rho\prod\limits_{\alpha=x,y} \frac{(1-2c^2_{i_\alpha})}{2^{c^2_{i_\alpha}}}\left[c^2_{i_\alpha} - 1 + c_{i_\alpha}u_\alpha + u^2_\alpha + \theta\right].
\end{equation}
When used within the OReg framework, the corresponding expressions for $\psi_\alpha$ and $\widetilde{\Pi}_{\alpha\beta}$ are obtained as:
\begin{widetext}
\vspace{-2em}
\begin{subequations}
    \begin{align}
		\psi_\alpha^{\text{OReg}|\text{GEq}}
		=
		&
        \frac{\tau\rho u_\alpha}{\theta}
        \Big(
            u_\alpha^2 + 3\theta - 1
        \Big)
        \bigg(
			\partial_\alpha u_\alpha - \frac{1}{2}\partial_\chi u_\chi
        \bigg)
        \sim
        \begin{cases}
            \mathcal{O}(u^5) \text{ if } \theta = 1/3
            \\
            \mathcal{O}(u^3) \text{ otherwise }
        \end{cases}
        \label{eq:guidedEq-oreg-mom-compatilbility}
        \\[-0.25em]
        \widetilde{\Pi}_{\alpha\beta}^{\text{OReg}|\text{GEq}}
        &=
        \frac{\rho \tau}{2\theta} \left(\partial_x u_x - \partial_y u_y \right)
        \begin{cases}
            -\left[\theta\left(1-3\theta\right) + u_x^2\left(u_x^2 - 1 \right)\right] &\text{if $\alpha=\beta=x$}\\
		    \textcolor{white}{-}\left[\theta\left(1-3\theta\right) + u_y^2\left(u_y^2 - 1 \right)\right] &\text{if $\alpha=\beta=y$}\\
		    \textcolor{white}{-[\theta(1-1]}u_x u_y (u_x^2 - u_y^2) &\text{if $\alpha\neq\beta$}.
		\end{cases}
		\label{eq:guidedEq-oreg-stress-tensor-error}
    \end{align}
    \vspace{-1em}
\end{subequations}
\end{widetext}
\noindent
Earlier work demonstrated that $\widetilde{\Pi}_{\alpha\beta}^{\text{OReg}|\text{GEq}}$ can be assimilated within $\Pi_{\alpha\beta}^\text{NS}$ and the errors from $\psi_\alpha^\text{OReg\textbar GEq}$ culmintate into an $\mathcal{O}(u^3)/(u)$ accurate, variable viscosity representation of NS hydrodynamics when operating at the lattice reference temperature $\theta_0 = 1/3$ and arbitrary temperatures respectively \citep{jonnalagaddaPRR2025}.
In this representation, the OReg-GEq stress tensor takes the following $\mathcal{O}(u^5)$ accurate variable viscosity form:
\vspace{-0.25em}
\begin{subequations}
    \begin{align}
        \displaystyle
        \Pi_{\alpha\beta}^{\text{OReg}|\text{GEq}}
        =
        &
        -\left(\tau\rho\theta\right)
        \mu^{\text{OReg}}_{\alpha\beta}
        \left(\partial_\alpha u_\beta + \partial_\beta u_\alpha - \partial_\chi u_\chi\delta_{\alpha\beta}\right) \nonumber
        \\
        &
        \hspace{2em}
        +
        \bar{\delta}_{\alpha\beta}\,
        \mathcal{O}(u^6)
        \label{eq:guidedEq-oreg-neq-pressure}
        \\[-1.25em]
        \intertext{where, \vspace{-1em}}
        \mu^{\text{OReg}}_{\alpha\beta}
        =
        &
        \begin{cases}
            \frac{1}{2\theta^2}(\theta - u_\alpha^2)(1-\theta-u_\alpha^2) &\mbox{ if } \alpha=\beta \\
            \phantom{\frac{1}{2\theta^2}(\theta - u_\alpha^2)} 1 &\mbox{ otherwise,}
        \end{cases}
        \label{eq:guidedEq-oreg-viscosity}
    \end{align}
\end{subequations}

\noindent
and $\bar{\delta}_{\alpha}$ is the complement of the Kronecker delta that vanishes when $\alpha = \beta$.
Here, evaluating partially corrected populations $\Psi_i^\text{P}$ that only eliminate $\psi_\alpha^{\text{OReg}|\text{GEq}}$ would yield a family of $\mathcal{O}(u^5)$ variable-viscosity kinetic models for the Navier-Stokes equations.
One possible set of such correction populations are:
\begin{equation}
\label{eq:partial-corrections}
\Psi_i^\text{P} =
    -\frac{1}{2}
	\begin{cases}
    c_{i_\alpha}
    \psi_\alpha
    &
    \begin{cases}
        \alpha = x \text{ if } i \in \left\lbrace 1, 3 \right\rbrace,\\
        \alpha = y \text{ if } i \in \left\lbrace 2, 4 \right\rbrace,\\
    \end{cases}
    \\
    0 &\text{otherwise.}
\end{cases}
\end{equation}
We drop the superscript $\text{OReg}|\text{GEq}$ for better readability and note that \Cref{eq:partial-corrections} does not require explicit evaluations of $\psi_\alpha$ through the analytical expression presented in \Cref{eq:guidedEq-oreg-mom-compatilbility}.
Indeed, since $\psi_\alpha = \displaystyle \sum\limits_i c_{i_\alpha} f_i^\text{OReg}$ is a moment of the known $f_i^\text{OReg}$ populations, obtaining fully-local partially corrected OReg models is a straightforward exercise.

Next, we present a realization of completely corrected populations, $\Psi_i^\text{C}$, that explicitly eliminate both $\psi_\alpha^{\text{OReg}|\text{GEq}}$ and $\widetilde{\Pi}_{\alpha\beta}^{\text{OReg}|\text{GEq}}$ as:
\begin{equation}
	\label{eq:complete-corrections}
    \Psi_i^\text{C}
    =
    -\frac{1}{4}
    \begin{cases}
     0
     \hspace{10em}
     \text{ if }  i = 0,\\
     -2\left( \widetilde{\Pi}_{\bar{\alpha}\bar{\alpha}} - c_{i_\alpha}\psi_\alpha \right)
     \hspace{1em}
     \begin{cases}
     	\alpha = x \text{ if } i \in  \left\lbrace 1, 3 \right\rbrace,\\
     	\alpha = y \text{ if } i \in  \left\lbrace 2, 4 \right\rbrace,\\
     \end{cases}
     \\
     \left(
     	\widetilde{\Pi}_{xx}
     	+
     	\widetilde{\Pi}_{yy}
     	+
     	c_{i_\alpha}
     	c_{i_{\bar{\alpha}}}
     	\widetilde{\Pi}_{xy} \right)
      \text{ otherwise. }
    \end{cases}
\end{equation}
where $\bar{\alpha} = y $ if $\alpha = x$ and vice-versa.
Once again, the superscript $\text{OReg}|\text{GEq}$ is dropped for better readability.
As shown in \Cref{eq:guidedEq-oreg-stress-tensor-error}, the stress tensor modelling errors $\widetilde{\Pi}_{\alpha\beta}$ appearing in \Cref{eq:complete-corrections} require velocity gradients to be evaluated.
Here, we leverage the interpretation of the OReg populations as one-step corrections to $f_i^{(neq)}$ \citep{jonnalagaddaPRR2025} to evaluate these gradients.
With such an interpretation, the viscous non-equilibrium thermodynamic forces $\mathrm{X}_{\alpha\beta}$ can be approximated locally via the first order accurate incompressible stress tensor as:
\begin{multline}
\label{eq:localderivatives}
\sigma  = \tau\rho\theta \left(\partial_\alpha u_\beta + \partial_\beta u_\alpha \right) = -\sum\limits_i c_{i_\alpha}c_{i_\beta}  f_i^\text{OReg}.
\end{multline}
Correspondingly, the pre-factor in \Cref{eq:guidedEq-oreg-stress-tensor-error} can be computed as:
\vspace{-0.5em}
\begin{equation}
\label{eq:geq-prefactor-evaluation}
\frac{\rho \tau}{2\theta} \left(\partial_x u_x - \partial_y u_y \right) = -\frac{1}{4\theta^2}\displaystyle\sum\limits_{i}(c_{i_x}^2 - c_{i_y}^{2})f_i^{\text{OReg\textbar GEq}}.
\vspace{-0.75em}
\end{equation}

The integration of the proposed correction models into an OReg LB timeloop is straightforward.
As shown in \Cref{fig:collision-workflow}, two additional steps evaluating $\Psi_i^\text{P/C}$ are required in the OReg collision kernel.
The rest of the LB algorithm proceeds without any change.

\begin{figure}
\scalebox{0.9}{
	\centering
    \begin{tikzpicture}
		\node[
			text width=7.6cm,
			align=center,
		] (corrected) {
			    \underline{\textbf{Evaluate correction populations}}

			    Compute $\displaystyle\psi_\alpha^{\text{OReg}|\text{GEq}} =
    \sum\limits_{i=0}^{8} c_{i_\alpha} f_i^{\text{OReg}|\text{GEq}}$
		};

		\node[
			draw,
			text width = 4.6cm,
			align=left,
			below =2.5pt of corrected,
			xshift = 2.1cm
		] (complete) {
		\vspace{-1em}
		\begin{center}
			\underline{\textbf{Complete}}

            \vspace{0.25em}
            Compute:
		\end{center}
		\vspace{-0.75em}
		(i)~Prefactor \big(\Cref{eq:geq-prefactor-evaluation}\big)

		(ii) $\widetilde{\Pi}_{\alpha\beta}^{\text{OReg}|\text{GEq}}$ \big(\Cref{eq:guidedEq-oreg-stress-tensor-error}\big)

		\vspace{0.25em}
		(iii) $\Psi_i^\text{C}$ \big(\Cref{eq:complete-corrections}\big)
		};

		\node[
			draw,
			text width = 3cm,
			align=left,
			left = 0.75cm  of complete,
		] (partial) {
		\vspace{-1em}
		\begin{center}
			\underline{\textbf{Partial}}

            Compute

            $\Psi_i^\text{P}$ \big(\Cref{eq:partial-corrections}\big)

		\end{center}
		};

		\node[
			draw,
			rectangle,
			fit=(corrected) (partial) (complete),
			inner sep=5pt,
			yshift=-5pt
		] (correctedBox) {};
		\node[
			draw,
			rectangle,
			text width=4.25cm,
            above = 0.5cm of correctedBox.north,
			align=left,
		] (base) {
		Compute:

		\vspace{0.25em}
		(i) $f_i^\text{GEq}$ \big(\Cref{eq:guidedEq}\big)

		(ii) $f_i^{\text{OReg}|\text{GEq}}$ \big(\Cref{eq:oreg}\big)
		};

		\node[
			draw,
			rectangle,
			text width=6cm,
			below = 0.5cm of correctedBox.south,
			align=center,
		] (goreg) {
		Compute generalized OReg non-equlibrium populations

		\vspace{0.5em}
		$f_i^{\text{GOReg}|\text{GEq}}$ = $f_i^{\text{OReg}|\text{GEq}}$ + $\Psi_i^\text{P}$/$\Psi_i^\text{C}$
		};

		\node[
			draw,
			rectangle,
			text width=6cm,
			below = 0.5cm of goreg.south,
			align=center,
		] (collision) {
		Perform regularized collision
		\vspace{-1em}
		$$f_i^* = f_i^\text{GEq} + \left(1-\frac{1}{\tau}\right)f_i^{\text{GOReg}|\text{GEq}}$$
		};

		\draw[-{Latex[length=5pt, width=5pt]}] (base) -- (correctedBox);
		\draw[-{Latex[length=5pt, width=5pt]}] (correctedBox) -- (goreg);
		\draw[-{Latex[length=5pt, width=5pt]}] (goreg) -- (collision);
	\end{tikzpicture}
    }
	\caption{\label{fig:collision-workflow}Collision operation for the partially corrected and completely corrected OReg\textbar GEq setup. In the uncorrected model, the regularized collision is performed directly with $f_i^\text{OReg\textbar GEq}$.}
\end{figure}

A few comments are in order before proceeding.
Firstly, we highlight that, while the corrected models presented in Equations (\ref{eq:partial-corrections}) and (\ref{eq:complete-corrections}) have complicated analytical representations that require additional computations, the overall increase in arithmetic intensity amounts to only approximately 11\% and 23\% respectively.
However, it is important to note that any performance penalty associated with strided memory accesses required to read data from nearest neighbours is completely eliminated.
Thus, the associated overhead in the overall performance can be expected to be minor.

The second pertains to the accuracy of the completely corrected populations presented in \Cref{eq:complete-corrections}.
It is important to note that the velocity gradients evaluated through Equations (\ref{eq:localderivatives}) and (\ref{eq:geq-prefactor-evaluation}) hold exactly only for incompressible flows and under conditions yielding $\widetilde{\Pi}_{\alpha\beta} = 0$ (see \Cref{eq:oreg-pressure-tensor}).
Specifically, in situations with appreciable velocity divergence or non-zero stress tensor modelling errors $\widetilde{\Pi}_{\alpha\beta}$, as in the present OReg\textbar GEq setting, the numerical accuracy of the required velocity gradients deteriorate.
Nevertheless, this loss of numerical accuracy can be recovered by employing more robust mechanisms to evaluate $\mathrm{X}_{\alpha\beta}$ and will be the subject of a seperate communication.
\vspace{-1em}
\section{Numerical Results}
\vspace{-1em}
\label{sec:results}
\begin{figure}[t]
		{
		\centering
		\includegraphics[scale=1]{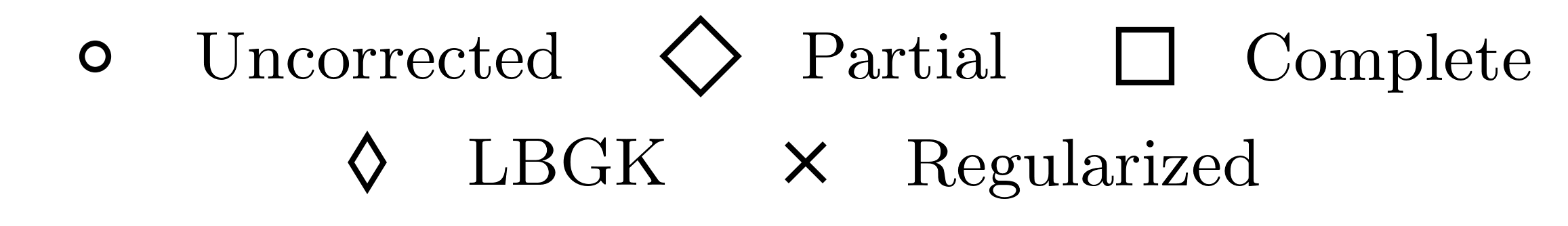}
		}

		\includegraphics[scale=1]{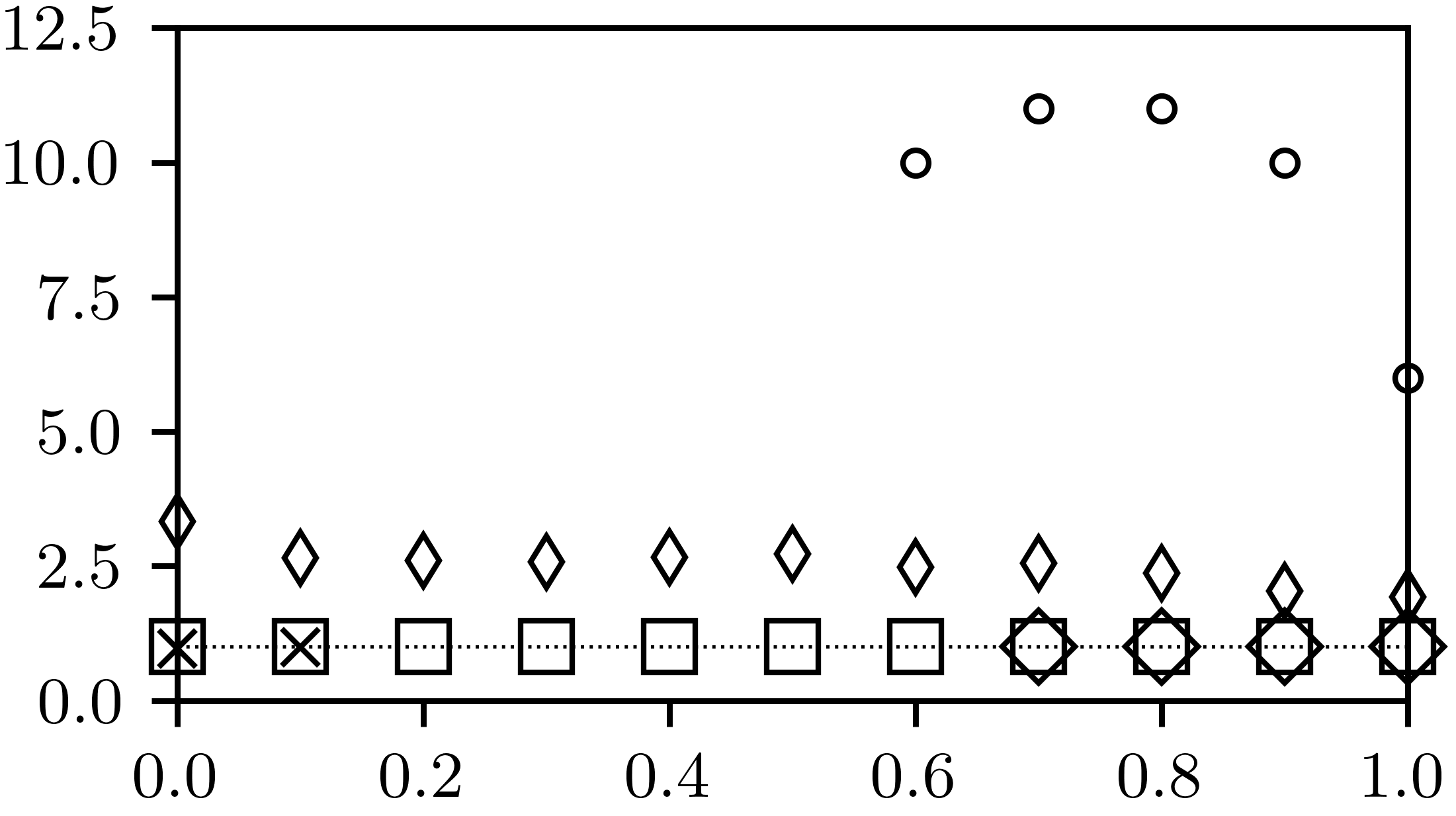}
		\begin{picture}(0,0)
			\put(-175,92.5){\footnotesize(a) $\theta = 0.25$}
			\put(-100,-4.5){\footnotesize Ma}
			\put(-215,52.5){ \rotatebox{90}{$\widetilde{\nu}/\nu$}}
		\end{picture}
		\vskip0.75em
		\includegraphics[scale=1]{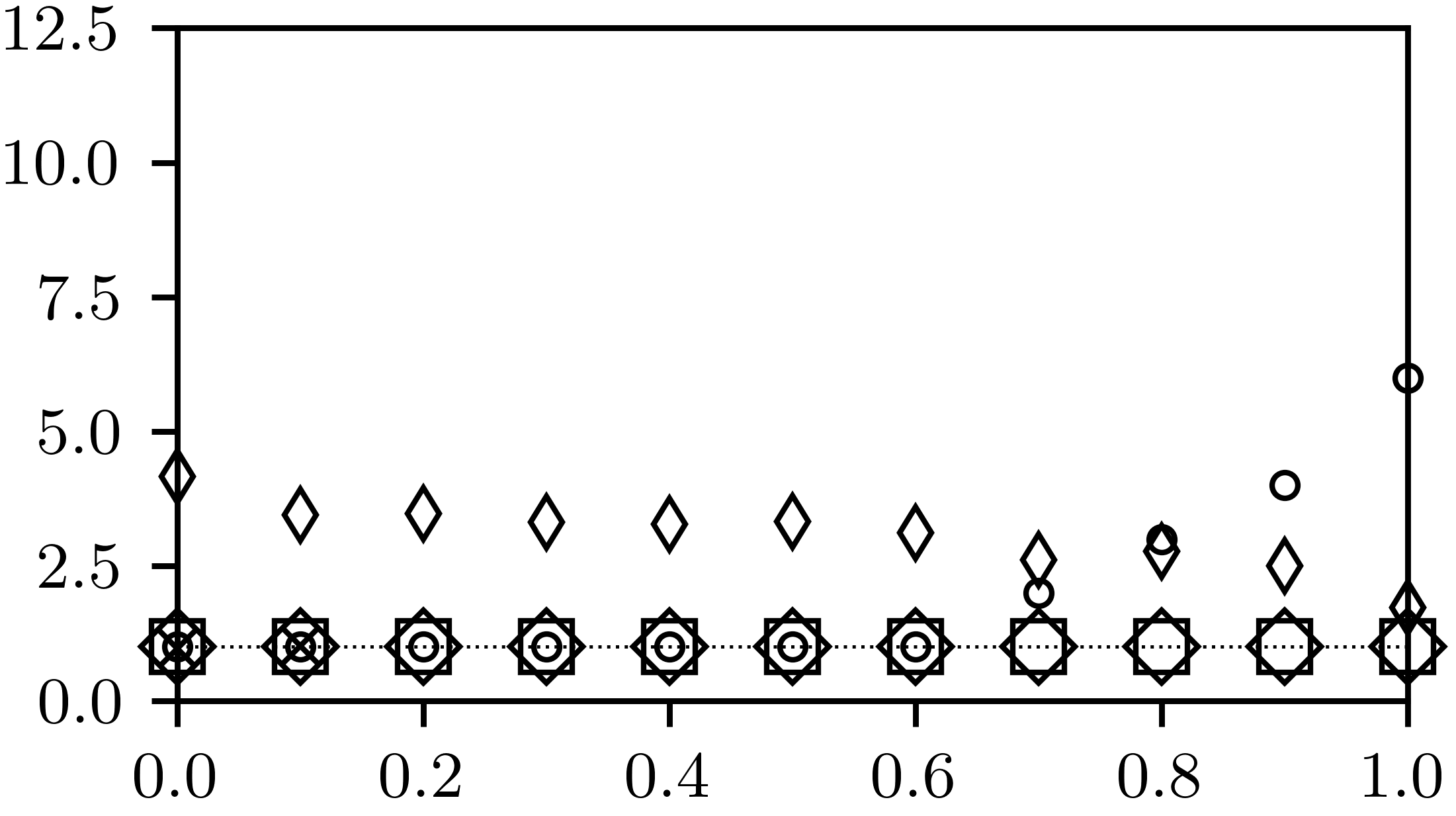}
		\begin{picture}(0,0)
			\put(-175,95){\footnotesize(b) $\theta = \theta_0 = 1/3$}
			\put(-100,-4.5){\footnotesize Ma}
			\put(-215,52.5){ \rotatebox{90}{$\widetilde{\nu}/\nu$}}
		\end{picture}
		\vskip0.75em
		\includegraphics[scale=1]{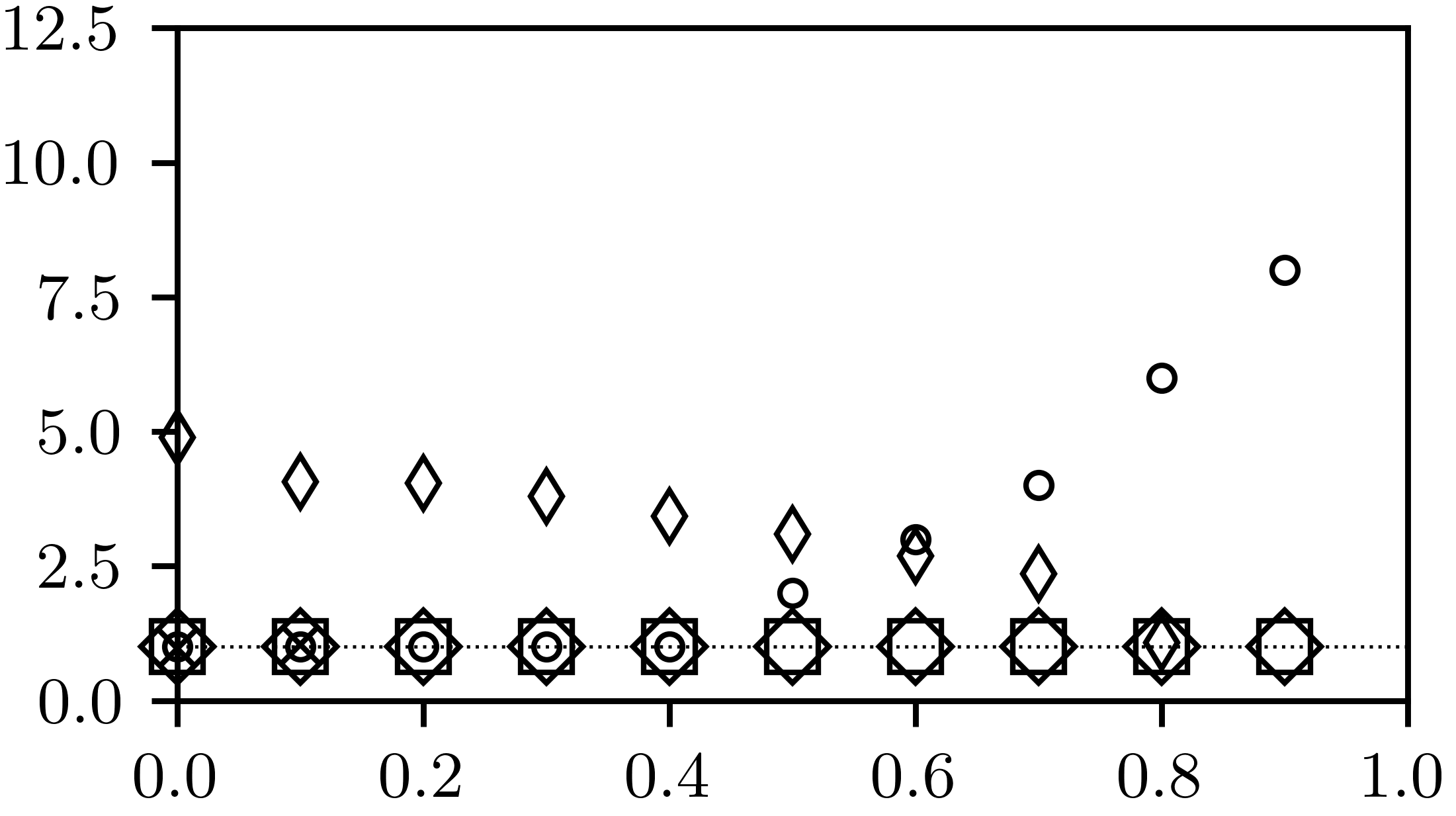}
		\begin{picture}(0,0)
			\put(-175,95){\footnotesize(c) $\theta = 0.4$}
			\put(-100,-4.5){\footnotesize Ma}
			\put(-215,52.5){ \rotatebox{90}{$\widetilde{\nu}/\nu$}}
		\end{picture}
		\vskip0.75em
		\includegraphics[scale=1]{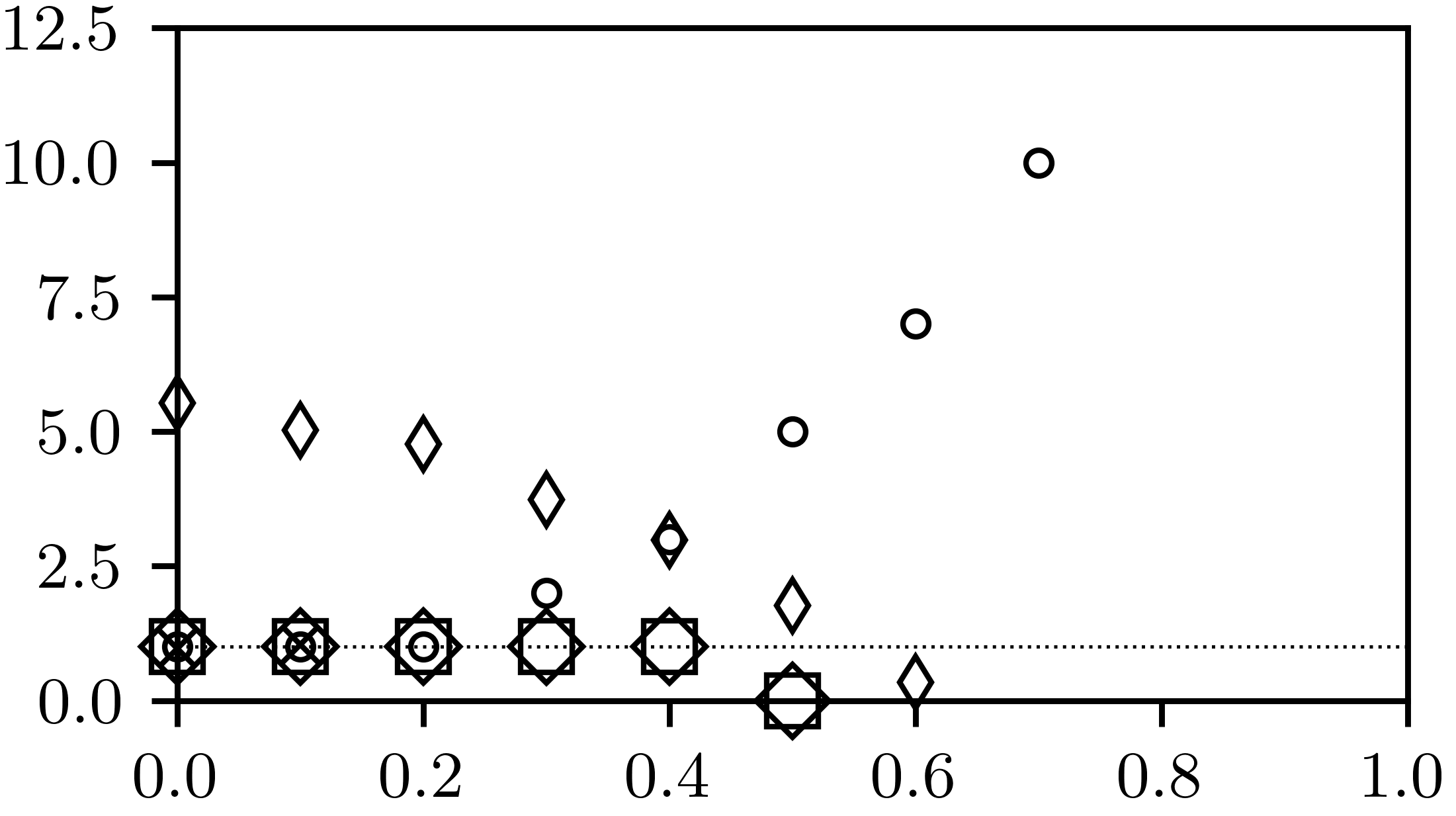}
		\begin{picture}(0,0)
			\put(-175,95){\footnotesize(d) $\theta = 0.5$}
			\put(-100,-4.5){\footnotesize Ma}
			\put(-215,52.5){ \rotatebox{90}{$\widetilde{\nu}/\nu$}}
		\end{picture}
		\caption{\label{fig:decaying-shear-wave}
		Deviation of the numerical viscosity ($\tilde{\nu}$) from the imposed viscosity ($\nu=0.001$) in a decaying rotated shear wave. Results are presented for the LBGK, regularized and the uncorrected, partially-corrected and, fully-corrected OReg schemes at different Mach numbers (Ma) and isothermal temperatures. Cases with catastrophic instabilities are not represented and the dotted line represents $\tilde{\nu}=\nu$.}
		\vskip-1em
	\end{figure}
\subsection{Decaying shear wave}
\vspace{-1em}
\sloppy
\noindent
We first consider the quasi-1D benchmark case of a fully-periodic decaying shear wave in which artificial viscous dissipation by numerical methods can be quantified.
Specifically, since observables decay at an exponential rate proportional to $\exp(-k_\alpha^2\widetilde{\nu}t)$, where $k_\alpha$ is the wave number and $\widetilde{\nu}$ is the numerical viscosity, a curve fitting exercise allows for the computation of $\widetilde{\nu}$.
Here, we consider a $\pi/4$ rotated decaying shear wave and evaluate the numerical behaviour of the uncorrected, partially and completely corrected OReg models.
The considered OReg schemes are compared against the LBGK and regularized LB scheme.

On a 1 $\times$ 200 spatial grid with a unit density and a lattice viscosity of $\nu=0.001$, we consider the wave to have initial velocity field as:
\begin{equation}
		u_x = u_y = \frac{\text{Ma}}{\sqrt{6}} + A_0 \sin\left[\frac{\sqrt{2}\pi}{N} (-x+y)\right],
\end{equation}
where Ma is the Mach number, $A_0 = 0.001$ is the amplitude of the wave, and $N$ is the spatial discretization in the $y-$direction.
Note that while the same test case has been previously used for examining the uncorrected OReg-GEq formulation \citep{jonnalagaddaPRR2025}, the configurations examined in this work are significantly more severe.
Specifically, in this study we consider both the $x-$ and $y-$velocity components to be inclined, larger Mach number ranges, and multiple lattice temperatures.
The numerical viscosity, $\widetilde{\nu}$, is computed as described in \citep{jonnalagaddaPRR2025} after 1200 iterations.
The deviation of $\widetilde{\nu}$ from $\nu$ is reported in \Cref{fig:decaying-shear-wave} for four isothermal temperatures in terms of the ratio $\widetilde{\nu}/\nu$.
Cases comprising catastrophic instabilities are not represented.

From \Cref{fig:decaying-shear-wave}(a), it can be seen that while the LBGK scheme is consistently stable when the lattice temperature is maintained at $\theta = 0.25$, the associated numerical viscosity is incorrect at all considered Mach numbers.
The uncorrected OReg scheme is not always stable, and, whenever stable, yields significantly larger values of numerical viscosity.
The partially corrected OReg scheme slightly improves on the stability range of the uncorrected model and yields the correct numerical viscosity wherever stable.
In contrast, the completely corrected OReg scheme yields accurate results in the entire range of considered Mach numbers.
In the remaining situations, as seen in Figures \ref{fig:decaying-shear-wave}(b-d), the LBGK scheme shows a clear downward trend with the deviation in $\nu$ being large for small Ma and vice-versa.
With the uncorrected OReg scheme, we see that the lattice viscosity is exactly recovered for a limited range of Mach numbers beyond which the deviation increases rapidly ultimately culminating into failure.
Note that the maximum attainable Ma becomes smaller (0.6, 0.4 and 0.2) with increasing lattice temperature ($\theta$ = 1/3, 0.4 and 0.5).
In contrast, both the partially and completely corrected OReg schemes yield accurate viscosity values upto Ma = 1, 0.9 and 0.4 for the three considered lattice temperatures.
Finally, for all cases with the regularized LB scheme, accurate viscosity values are only obtained in the nearly incompressible limit with simulations at Ma $>$ 0.1 yielding unstable results.
\vspace{-1em}
\subsection{Isothermal shocktube}
\vspace{-1em}
\noindent
The uncorrected OReg-GEq formulation has also been evaluated previously using the quasi-1D isothermal shocktube benchmark case.
Specifically, while it yielded accurate results without any spurious oscillations  for two different configurations of lattice temperatures and viscosities, ($\theta, \nu$) = (0.35,$10^{-5}$) and (0.4,$10^{-9}$) respectively, the transition from the high-density to the equilibrated regions was found to be longer due to first-order accuracy at $\theta\neq 1/3$ \citep{jonnalagaddaPRR2025}.
Here, we employ an 800$\times$1 grid with fullway bounce-back wall representations and consider two experiments to study the behaviour of the partially and completely corrected OReg schemes against the analytical solution \citep{athermalshocktube_analytical_sol} as well as the LBGK, regularized and uncorrected UC-OReg schemes.
The shocktube is initialized with a stationary fluid having a discontinuous density field of $\rho = 1.5$ for $x \leqslant 0.5$ and $\rho = 1.0$ for $x>0.5$.
All simulations are run for 400 time steps.
\begin{figure}
	\setlength{\belowcaptionskip}{-1.25em}
	\includegraphics[scale=1]{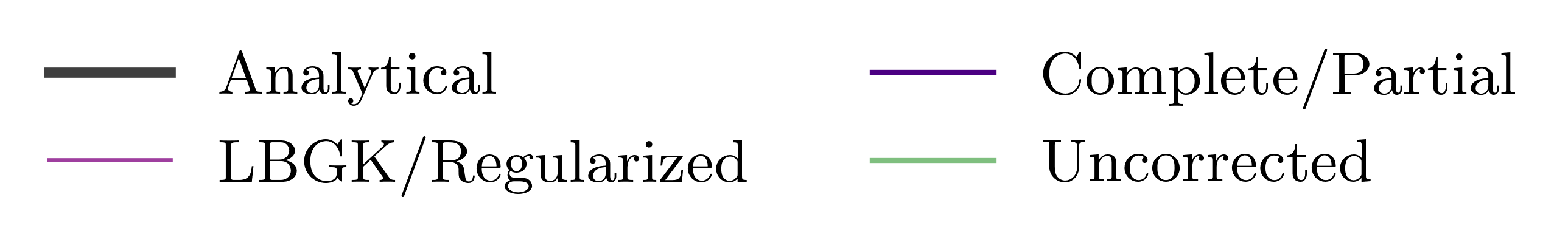}

	\vspace{0.5em}

	\includegraphics[scale=1]{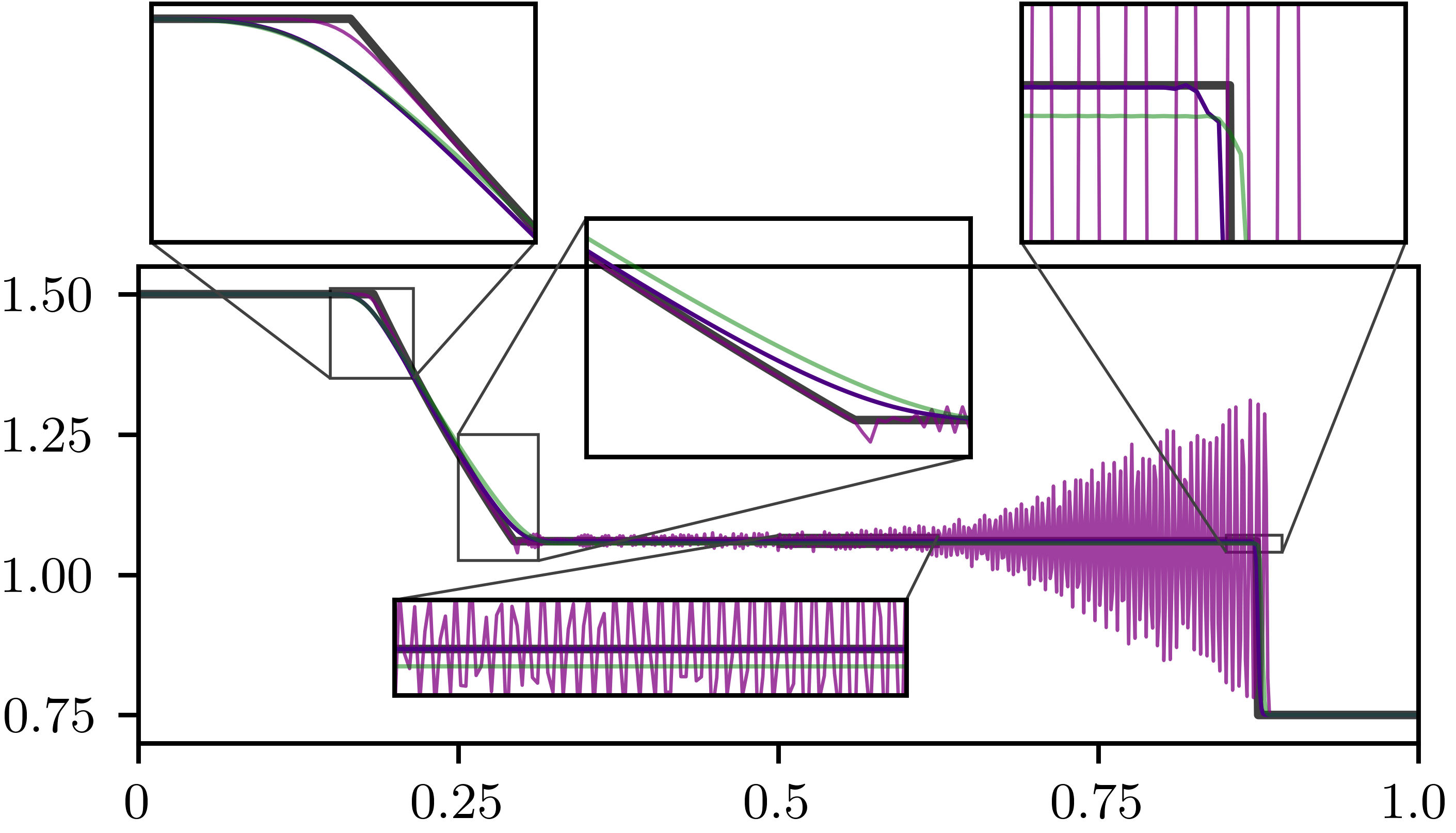}

	\includegraphics[scale=1]{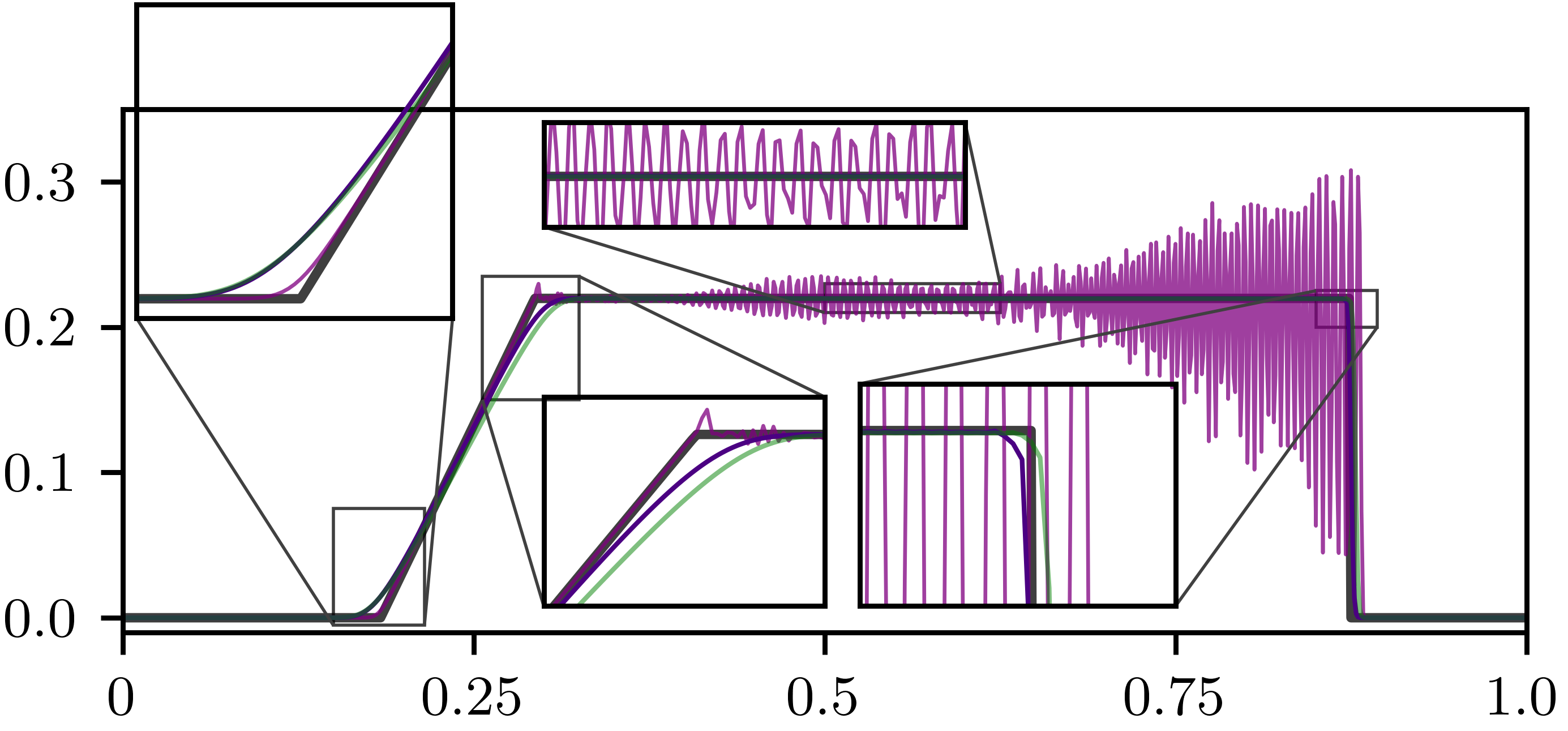}

	\begin{picture}(0,0)
            \put(-95,235){\footnotesize (i)}
            \put(-19,198){\footnotesize (ii)}
            \put(-100,118){\footnotesize (iii)}
            \put(-35,55){\footnotesize (iv)}
            \put(-67,165){\footnotesize (v)}
            \put(-50,100){\footnotesize (vi)}
            \put(106,235){\footnotesize (vii)}
            \put(69,30){\footnotesize (viii)}
            \put(-131.5,170){\rotatebox{90}{\footnotesize Density}}
            \put(-131.5,55){\rotatebox{90}{\footnotesize Velocity}}
            \put(10.75,123){\footnotesize $x$}
            \put(6,5){\footnotesize $x$}
        \end{picture}

     \vspace{-1em}
	\caption{\label{fig:ist-c1} Isothermal shocktube results for different LB schemes at a mildly elevated isothermal lattice temperatures ($\theta$ = 0.4) and an extremely small lattice viscosity of $\nu = 1 \times 10^{-12}$.}
\end{figure}

In the first experiment, while the lattice temperature is maintained at $\theta = 0.4$ as in the previous study, the lattice viscosity is reduced by three decades to $\nu = 10^{-12}$.
From \Cref{fig:ist-c1}, it can be seen that the LBGK and regularized LB schemes yield identical results and demonstrate significant numerical oscillations while the uncorrected OReg scheme is stable and free from spurious oscillations even at $\nu = 10^{-12}$ due to its variable viscosity nature.
However, the uncorrected OReg scheme still retains the incorrect slope in the transition region as shown in Figures \ref{fig:ist-c1}(i-iv).
Further, in comparison to the analytical solution, the uncorrected OReg scheme recovers a slightly smaller equilibrated density (\Cref{fig:ist-c1}(v)) while the equilibrated velocity (\Cref{fig:ist-c1}(vi)) is exactly recovered.
Finally, in the low-density region, Figures \ref{fig:ist-c1}(vii, viii) reveal that the equilibrium region is slightly longer than expected.
In contrast, both the partially and fully corrected OReg schemes accurately recover the density and velocity fields in the entire flow domain.

\begin{figure}
	\setlength{\belowcaptionskip}{-1em}
	\includegraphics[scale=1]{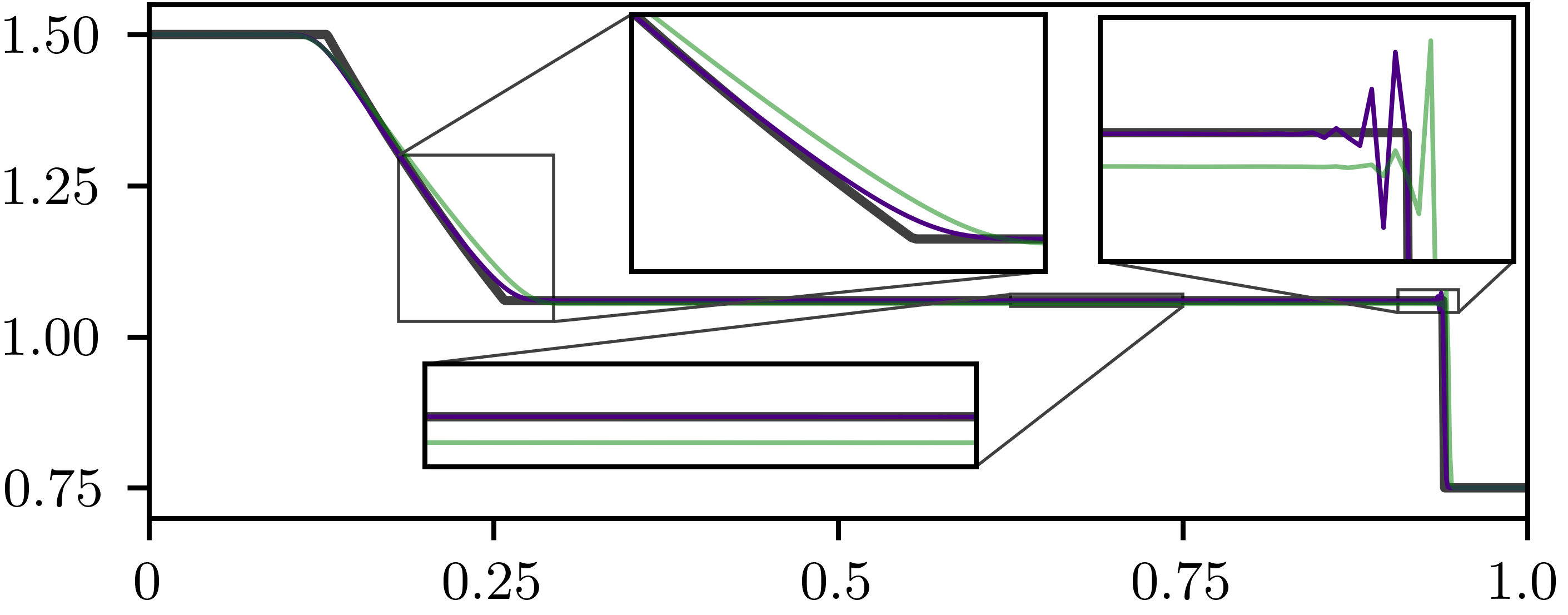}

	\vspace{1em}

	\includegraphics[scale=1]{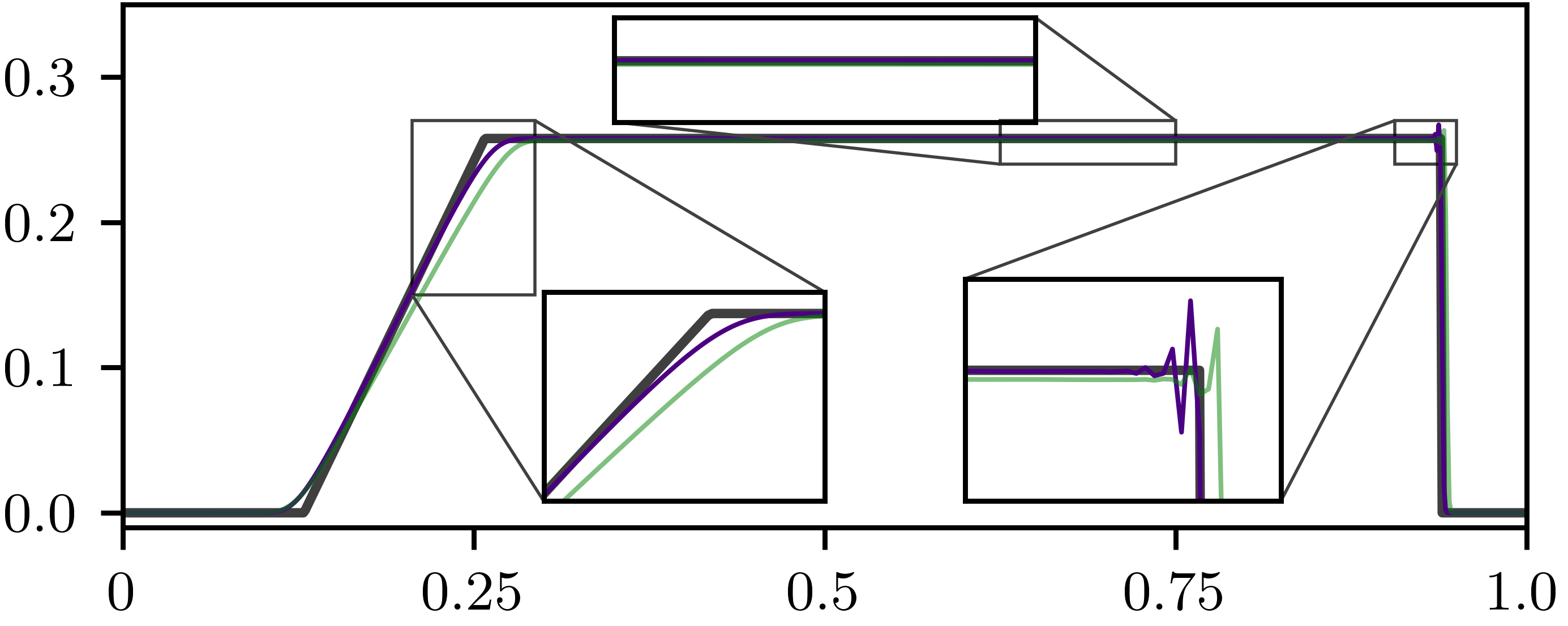}

	\begin{picture}(0,0)
            \put(-20,180){\footnotesize (i)}
            \put(-35.5,55){\footnotesize (ii)}
            \put(-69,155){\footnotesize (iii)}
            \put(-40,102){\footnotesize (iv)}
            \put(56.5,210){\footnotesize (v)}
            \put(32,58){\footnotesize (vi)}
            \put(-131.5,162){\rotatebox{90}{\footnotesize Density}}
            \put(-131.5,48){\rotatebox{90}{\footnotesize Velocity}}
            \put(10.75,116){\footnotesize $x$}
            \put(6,5){\footnotesize $x$}
        \end{picture}

	\vspace{-1em}
	\caption{\label{fig:ist-c2} Isothermal shocktube results for different LB schemes at an elevated isothermal lattice temperatures ($\theta$ = 0.55) and lattice viscosity of $\nu = 1 \times 10^{-9}$. The represented curves have the same meaning as that of \Cref{fig:ist-c1}. Note that the LBGK yield unphysical results for this case and is therefore not represented.}
	\vspace{-1.25em}
\end{figure}

\Cref{fig:ist-c2} presents the results of the second experiment which, as in the previous study, maintains the lattice viscosity at $\nu = 10^{-9}$ but employs a higher lattice temperature of $\theta = 0.55$.
It can be seen that the uncorrected OReg scheme, while stable, still retains all the deficiencies observed in the first experiment.
In contrast, and like the previous case, both the partially and completely corrected OReg schemes address all deficiencies.
We highlight that while the OReg schemes display numerical oscillations as shown in \Cref{fig:ist-c2}(v, vi), the magnitude of these artefacts are negligible even at temperatures as high as $\theta = 0.55$.
Conversely, the LBGK and regularized LB schemes yield unphysical results and, therefore, does not feature in \Cref{fig:ist-c2}.
\vspace{-1em}
\subsection{Doubly Periodic Shear Layer}
\begin{figure}
	\setlength{\belowcaptionskip}{-2em}
	\centering
	\includegraphics[scale=0.106]{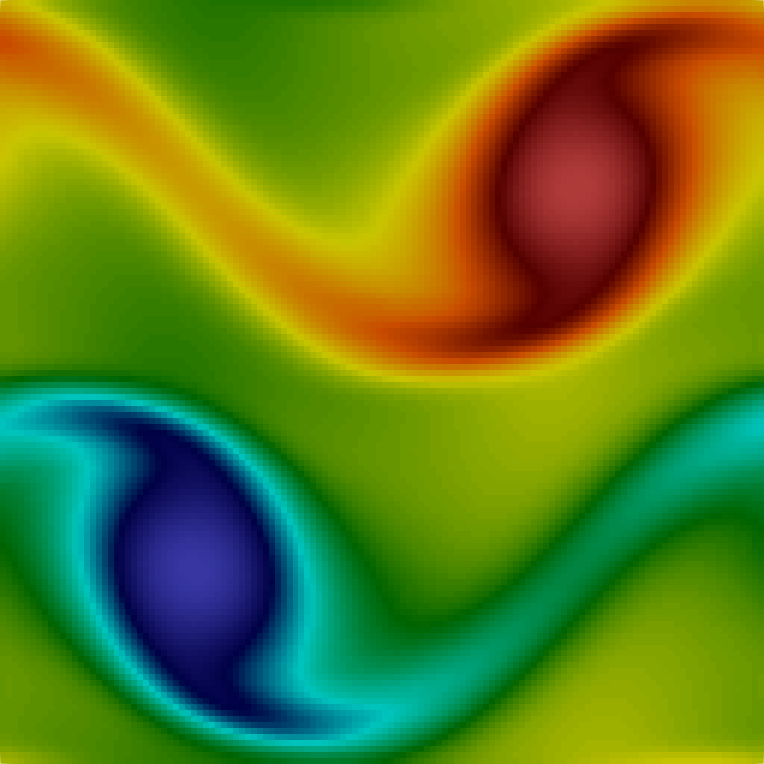}
	\begin{picture}(-4,0)
		\put(-20,5){\footnotesize(a)}
	\end{picture}
	\includegraphics[scale=0.106]{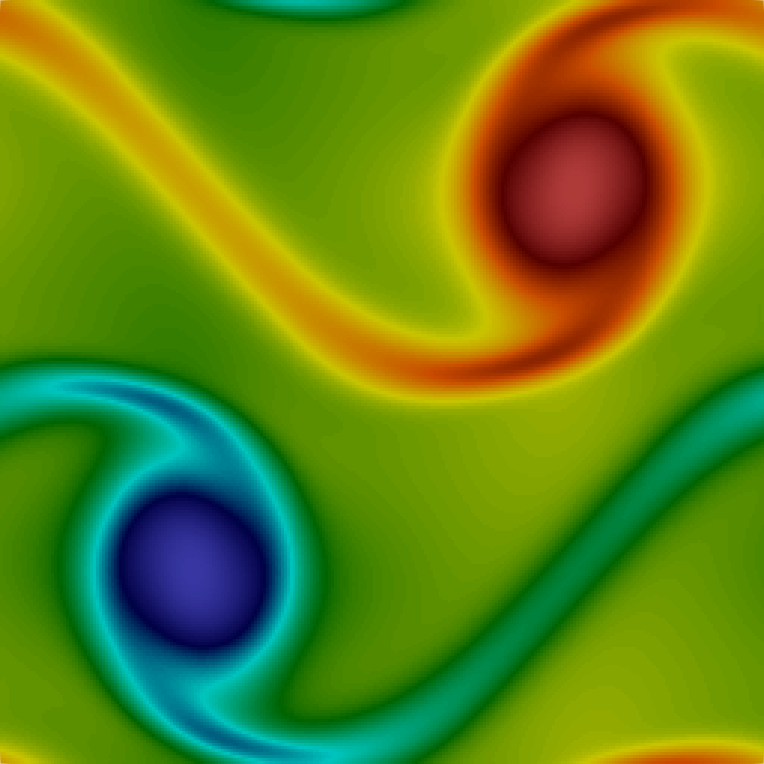}
	\begin{picture}(-4, 0)
		\put(-20,5){\footnotesize(d)}
	\end{picture}
	\includegraphics[scale=0.106]{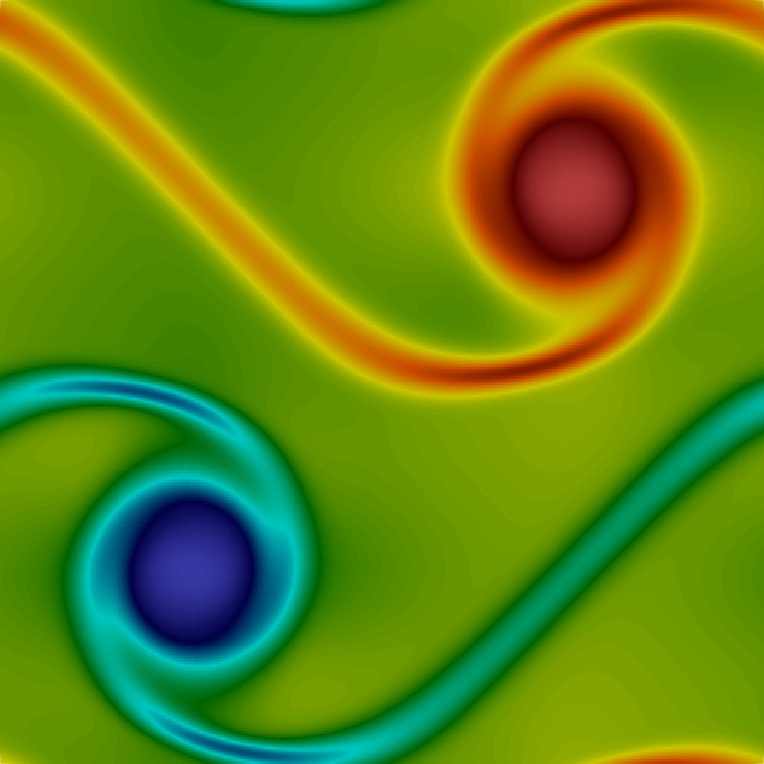}
	\begin{picture}(-4, 0)
		\put(-20,5){\footnotesize(g)}
	\end{picture}

	\vspace{0.1em}

	\includegraphics[scale=0.106]{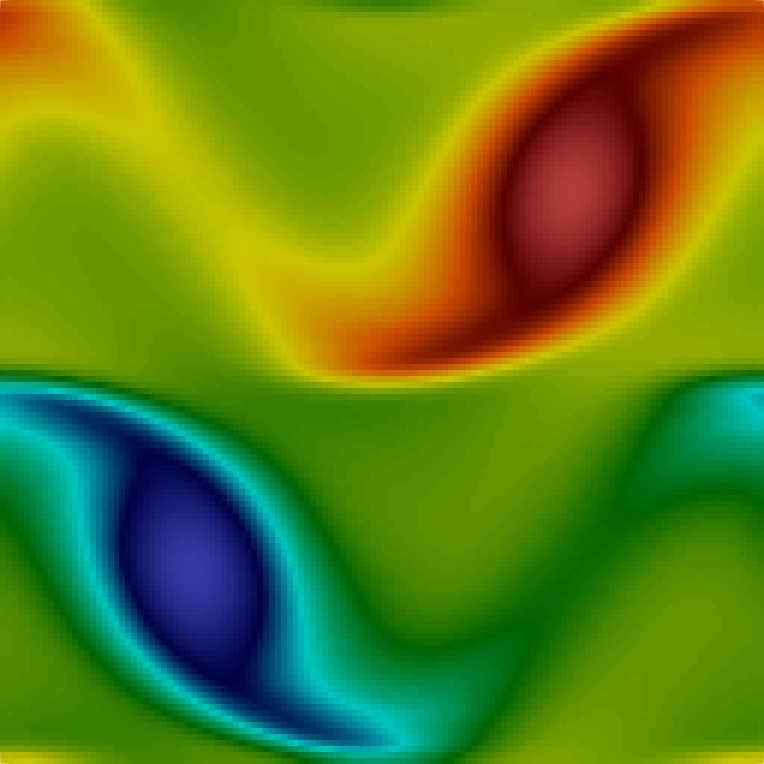}
	\begin{picture}(-4, 0)
		\put(-20,5){\footnotesize(b)}
	\end{picture}
	\includegraphics[scale=0.106]{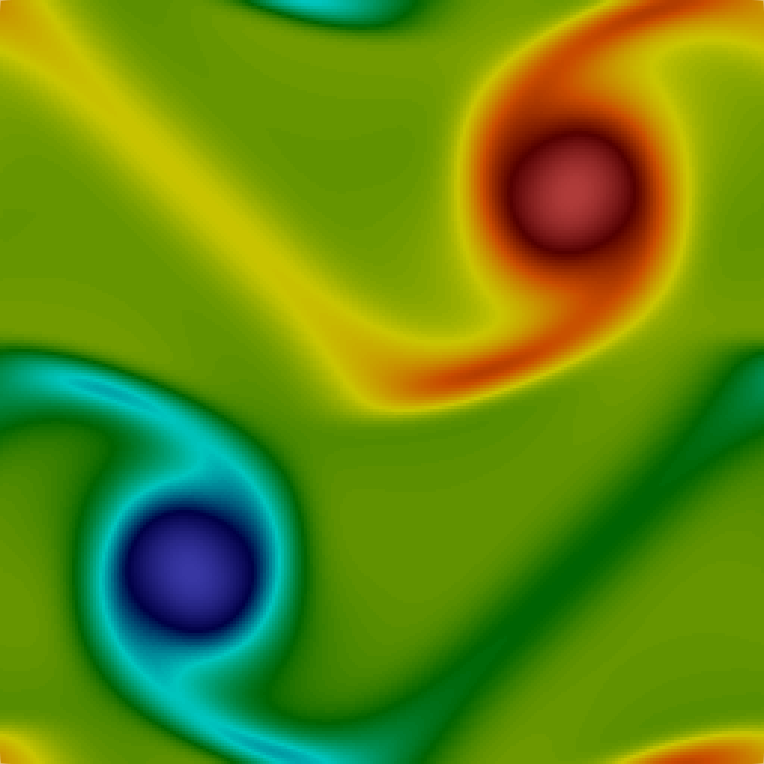}
	\begin{picture}(-4, 0)
		\put(-20,5){\footnotesize(e)}
	\end{picture}
	\includegraphics[scale=0.106]{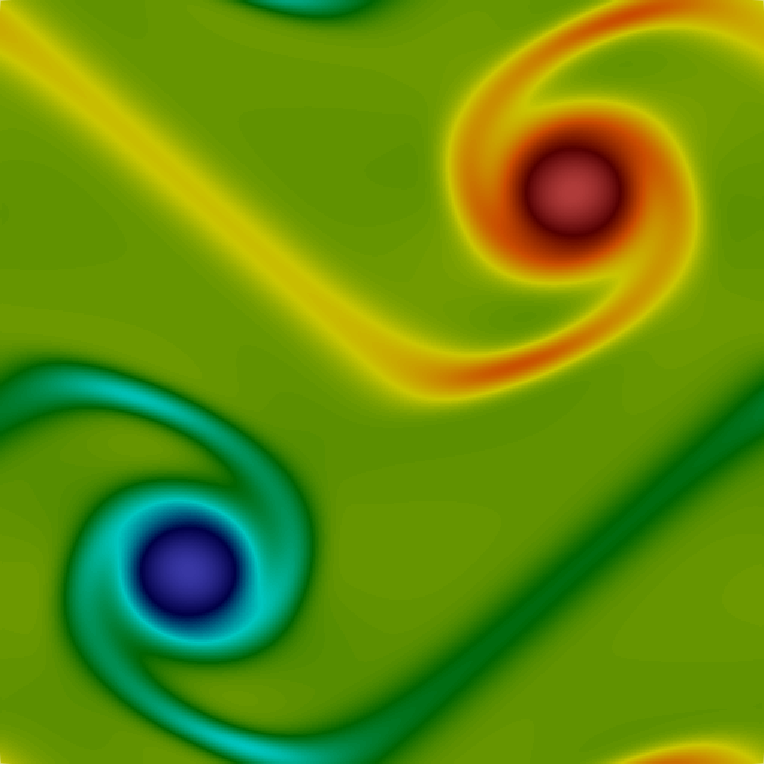}
	\begin{picture}(-4, 0)
		\put(-20,5){\footnotesize(h)}
	\end{picture}

	\vspace{0.1em}

	\includegraphics[scale=0.106]{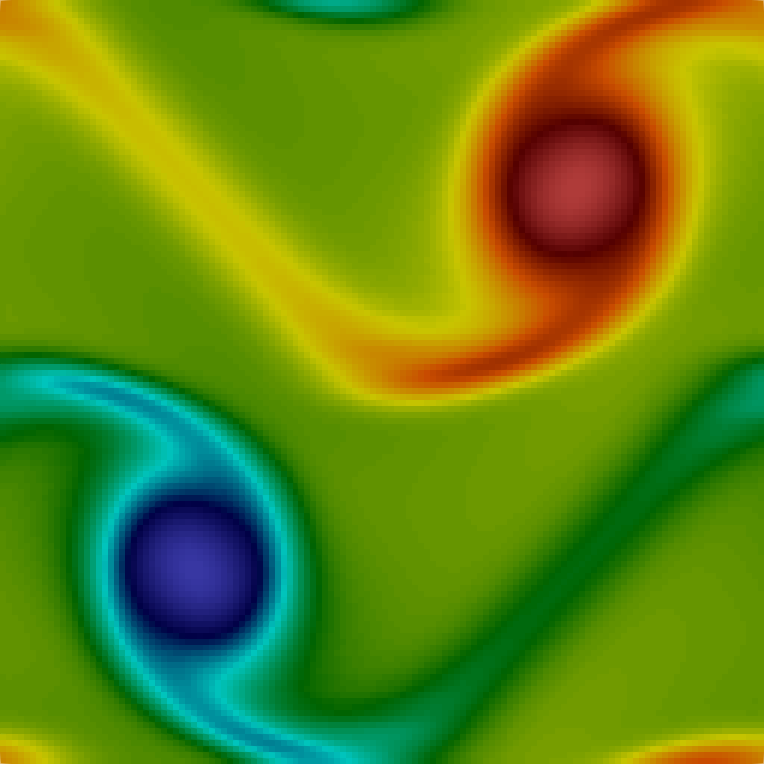}
	\begin{picture}(-4, 0)
		\put(-20,5){\footnotesize(c)}
	\end{picture}
	\includegraphics[scale=0.106]{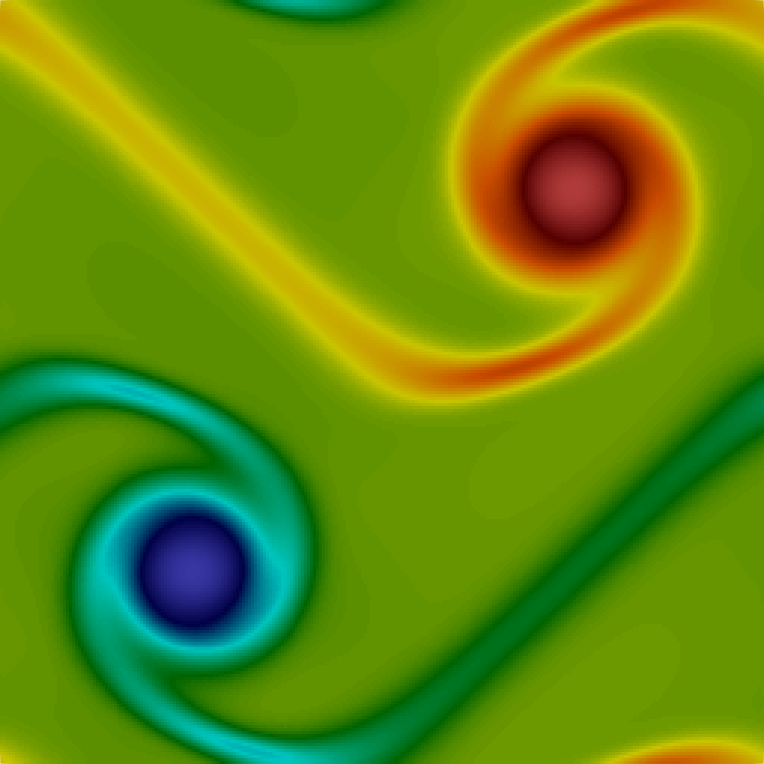}
	\begin{picture}(-4, 0)
		\put(-20,5){\footnotesize(f)}
	\end{picture}
	\includegraphics[scale=0.106]{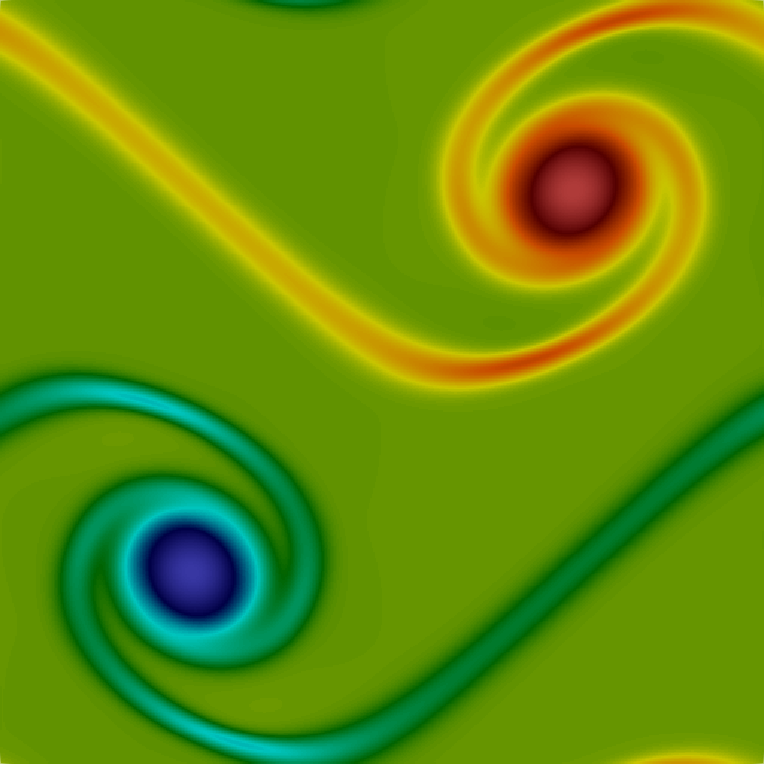}
	\begin{picture}(-4, 0)
		\put(-20,5){\footnotesize(i)}
	\end{picture}

	\caption{Normalized vorticity contours for the double periodic shear layer problem after unit characteristic time at an isothermal lattice temperature of $\theta=0.6$. Results are presented for the uncorrected, partially-corrected and completely-corrected OReg schemes (top to bottom) on $N^2$ sized grids with $N$ = 128, 256 and 512 (left to right). Simulations conducted with the Lattice-BGK and regularized LB schemes are unstable.\label{fig:dps-tp6-vorticity}}
	\vspace{1em}
\end{figure}

\vspace{-1em}
\noindent
The behaviours of both the proposed corrected OReg schemes have been indistinguishable from one another due to the quasi-1D nature of the two problems considered in the previous sections.
Thus, in this section, we consider a purely two dimensional non-linear problem namely the doubly-periodic shear layer benchmark case.
The $2\pi \times 2\pi$ simulation domain, discretized using $N$ grid points in each direction, is initialized with unit density and a velocity field given as:

\begin{subequations}
\begin{align}
		u_x &= 
		\begin{cases}
			u_0 \tanh \left[ \kappa \left(\frac{y}{N} - \frac{1}{4} \right) \right]\, , y \leq \frac{N}{2}\\
			u_0 \tanh \left[ \kappa \left(\frac{3}{4} - \frac{y}{N} \right) \right]\, , y > \frac{N}{2}\\
		\end{cases}\\
		\hspace{3em}& u_y = \delta u_0 \sin \left[2\pi\left(\frac{x}{N} + \frac{1}{4}\right) \right].
\end{align}
\end{subequations}
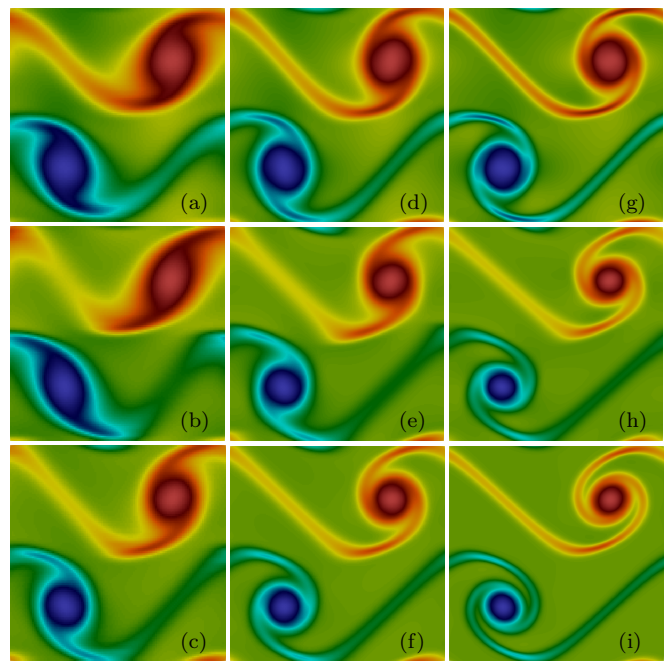
\begin{figure}
	\setlength{\belowcaptionskip}{-1em}
	\centering
    \hfill
	\includegraphics[scale=0.106]{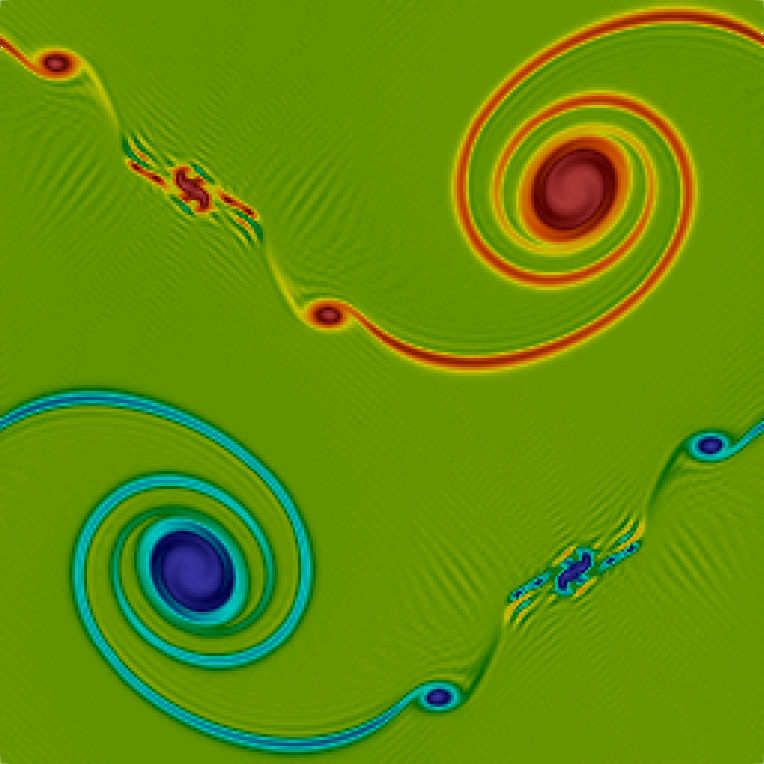}
	\begin{picture}(-4, 0)
		\put(-145,40){\footnotesize Unstable}
		\put(-100,5){\footnotesize(a)}
		\put(-20,5){\footnotesize(f)}
	\end{picture}
	\includegraphics[scale=0.106]{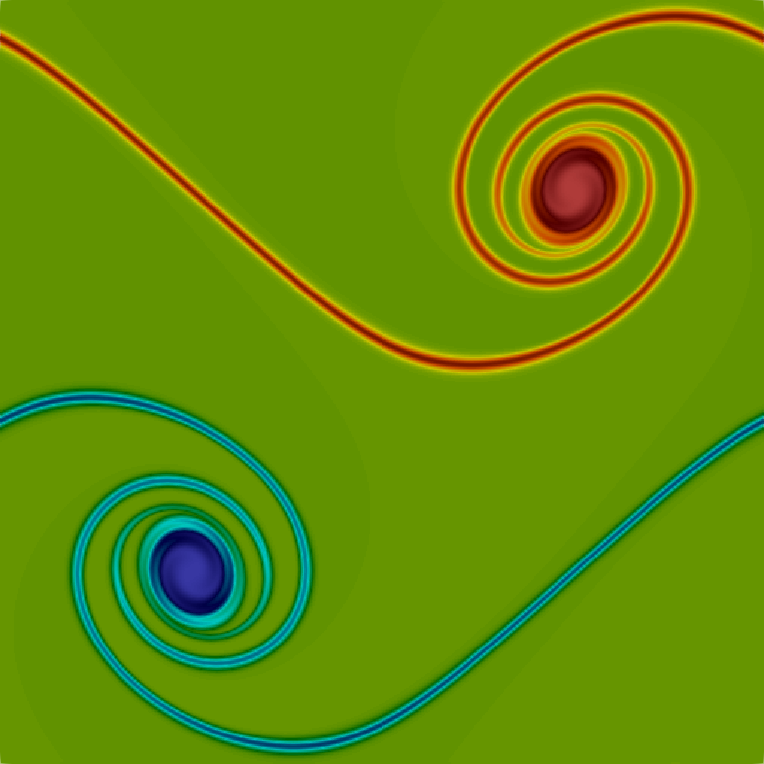}
	\begin{picture}(-4, 0)
		\put(-20,5){\footnotesize(k)}
	\end{picture}

	\vspace{0.1em}

	\includegraphics[scale=0.106]{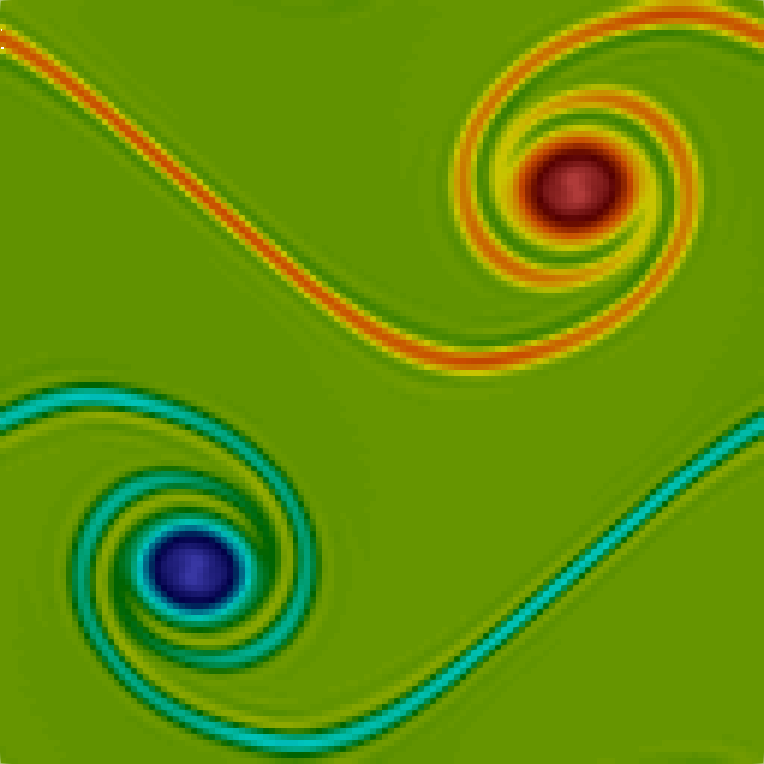}
	\begin{picture}(-4,0)
		\put(-20,5){\footnotesize(b)}
	\end{picture}
	\includegraphics[scale=0.106]{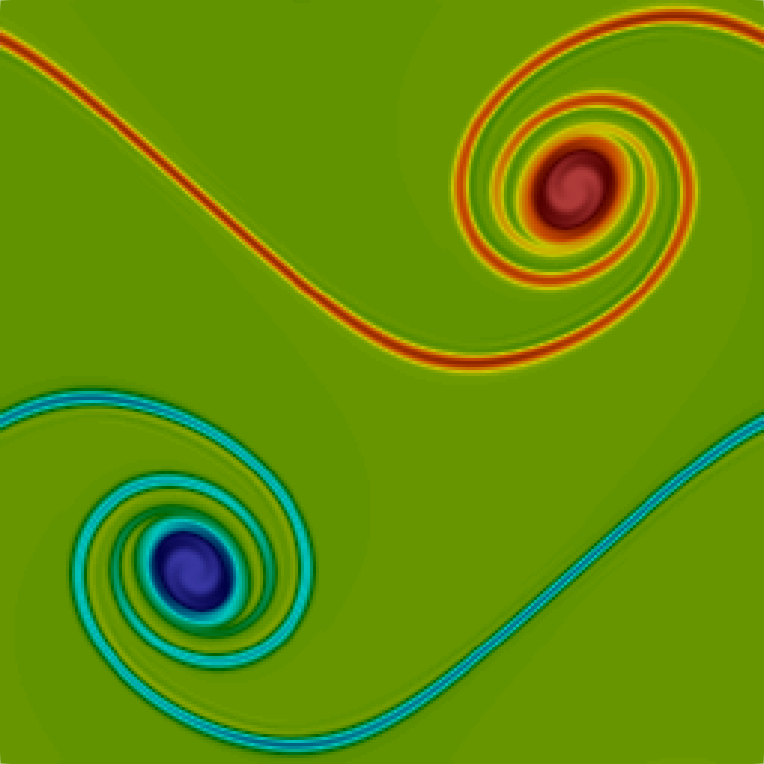}
	\begin{picture}(-4, 0)
		\put(-20,5){\footnotesize(g)}
	\end{picture}
	\includegraphics[scale=0.106]{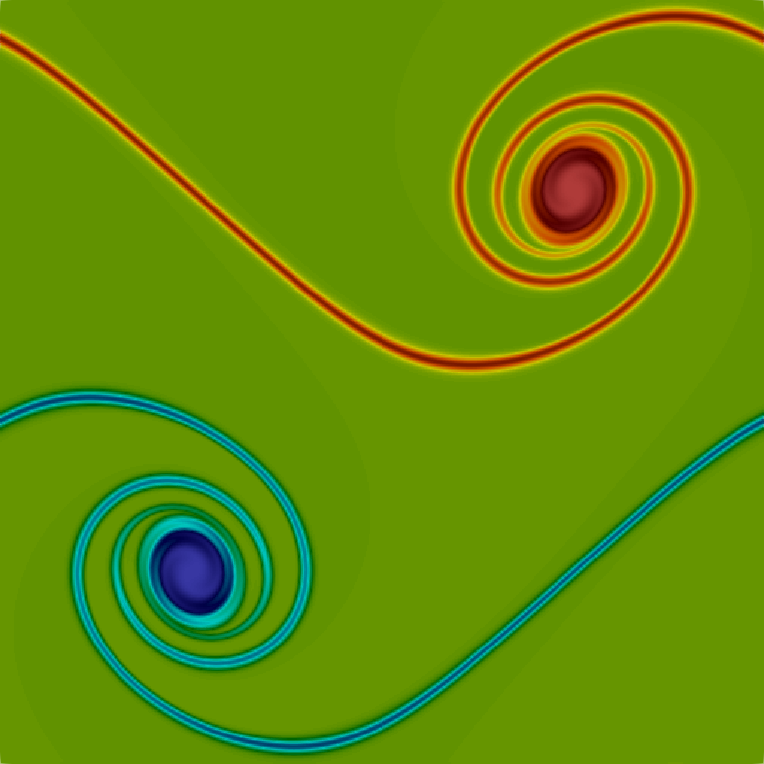}
	\begin{picture}(-4, 0)
		\put(-20,5){\footnotesize(l)}
	\end{picture}

	\vspace{0.1em}

	\includegraphics[scale=0.106]{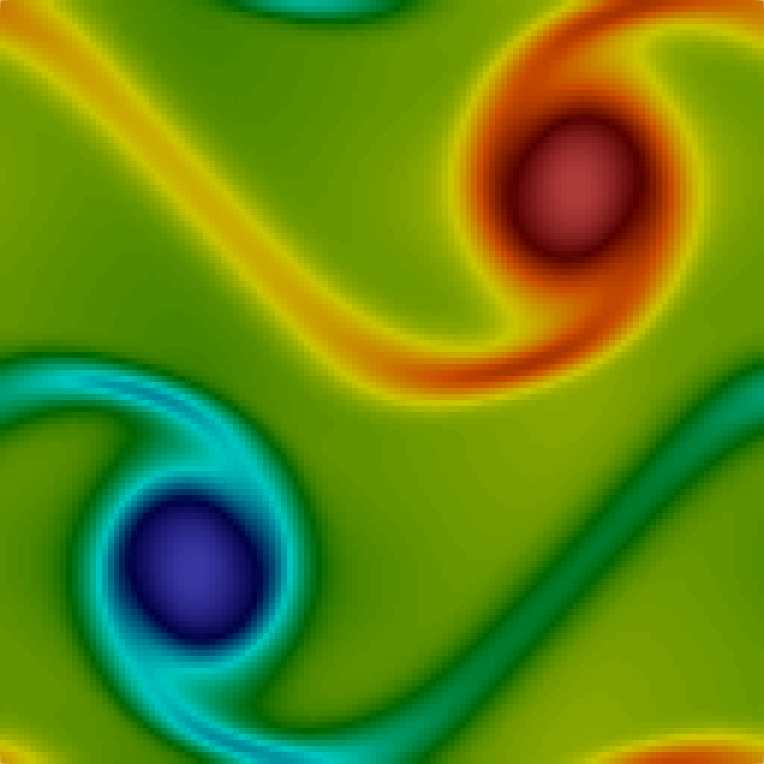}
	\begin{picture}(-4,0)
		\put(-20,5){\footnotesize(c)}
	\end{picture}
	\includegraphics[scale=0.106]{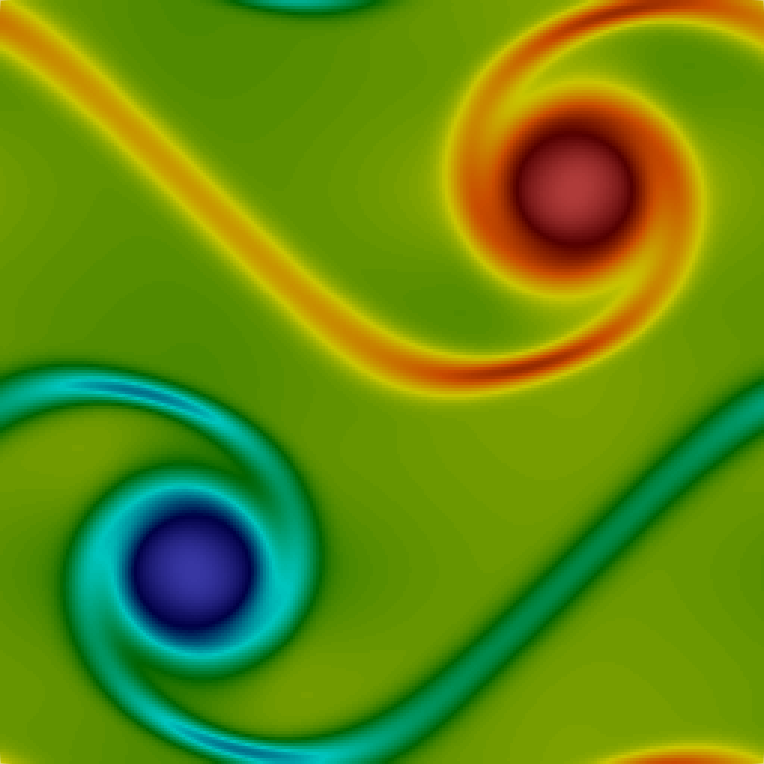}
	\begin{picture}(-4, 0)
		\put(-20,5){\footnotesize(h)}
	\end{picture}
	\includegraphics[scale=0.106]{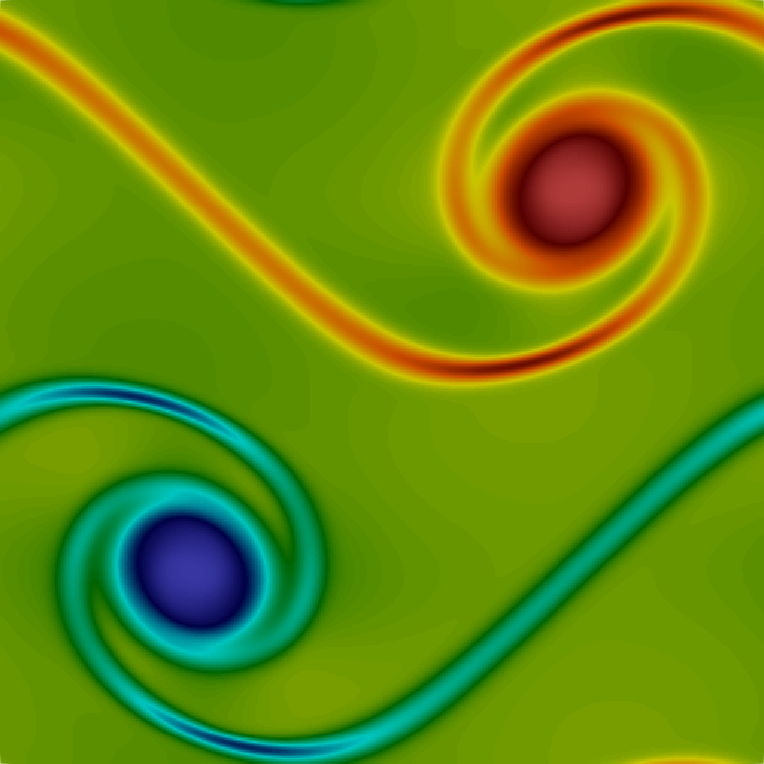}
	\begin{picture}(-4, 0)
		\put(-20,5){\footnotesize(m)}
	\end{picture}

	\vspace{0.1em}

	\includegraphics[scale=0.106]{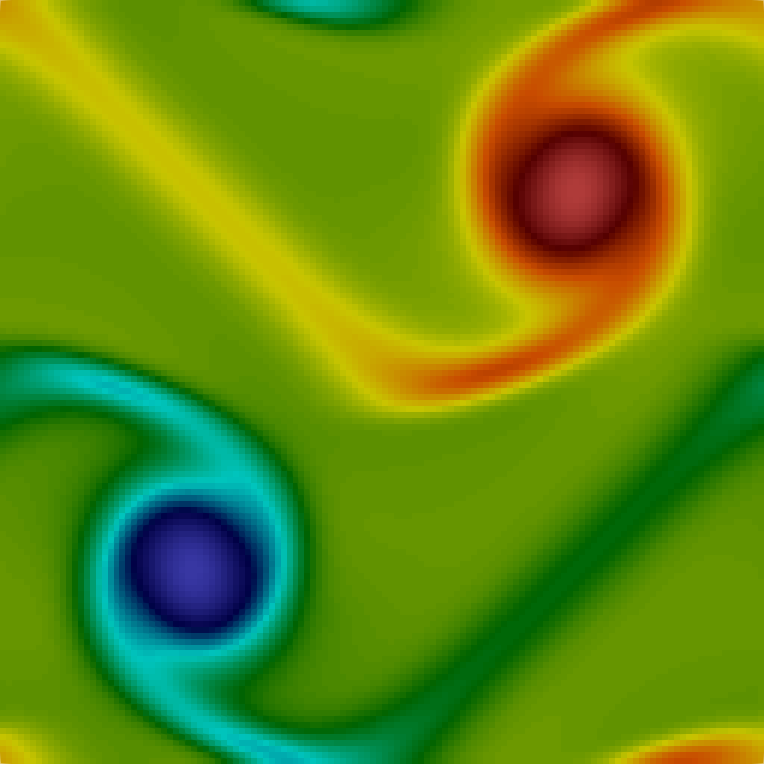}
	\begin{picture}(-4,0)
		\put(-20,5){\footnotesize(d)}
	\end{picture}
	\includegraphics[scale=0.106]{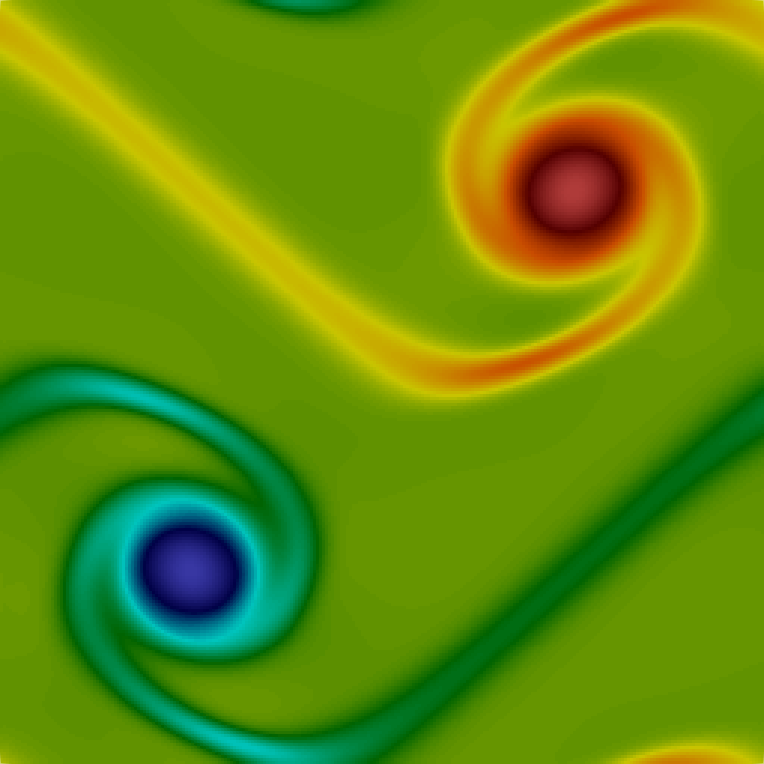}
	\begin{picture}(-4, 0)
		\put(-20,5){\footnotesize(i)}
	\end{picture}
	\includegraphics[scale=0.106]{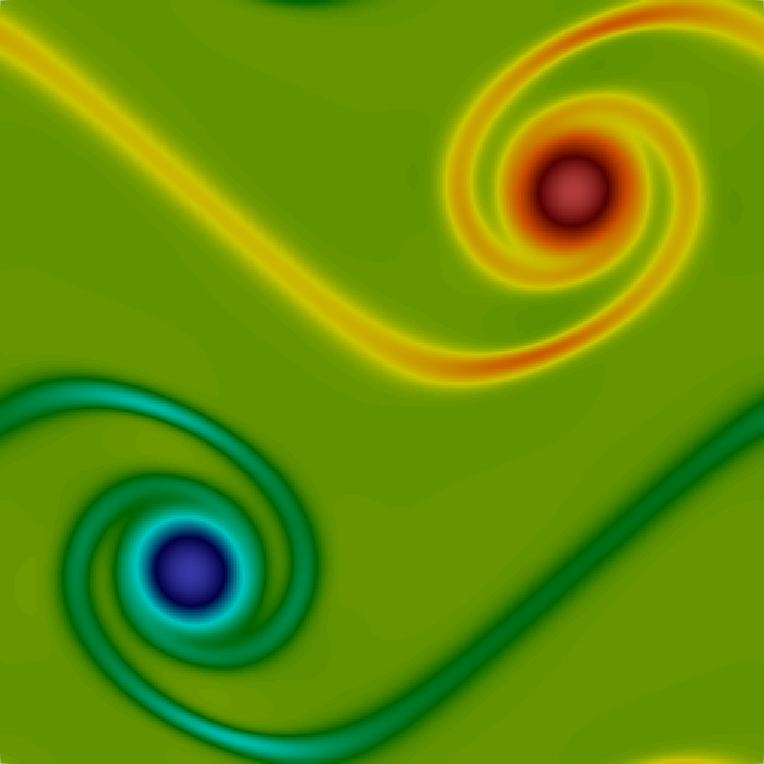}
	\begin{picture}(-4, 0)
		\put(-20,5){\footnotesize(n)}
	\end{picture}

	\vspace{0.1em}

	\includegraphics[scale=0.106]{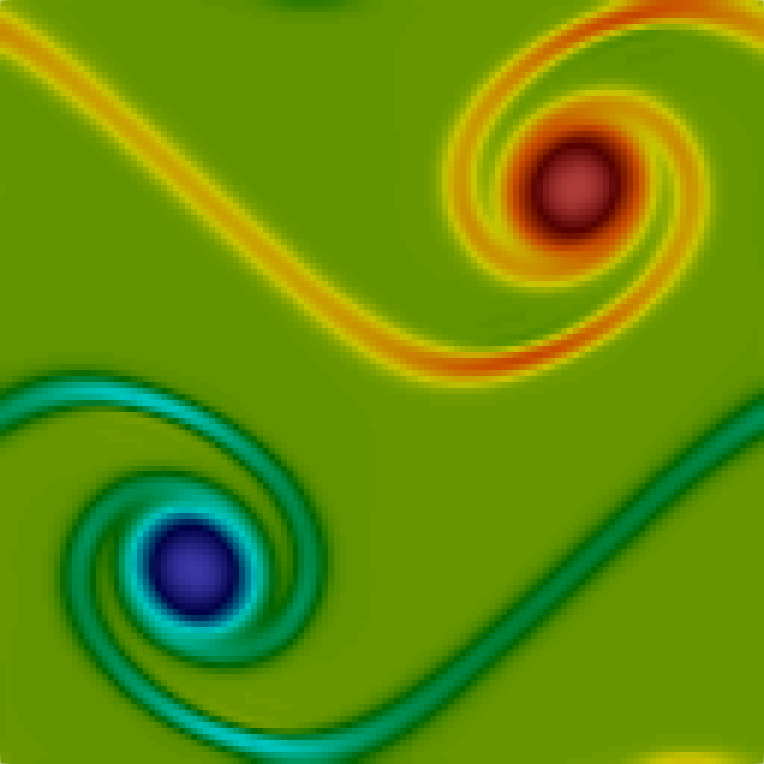}
	\begin{picture}(-4,0)
		\put(-20,5){\footnotesize(e)}
	\end{picture}
	\includegraphics[scale=0.106]{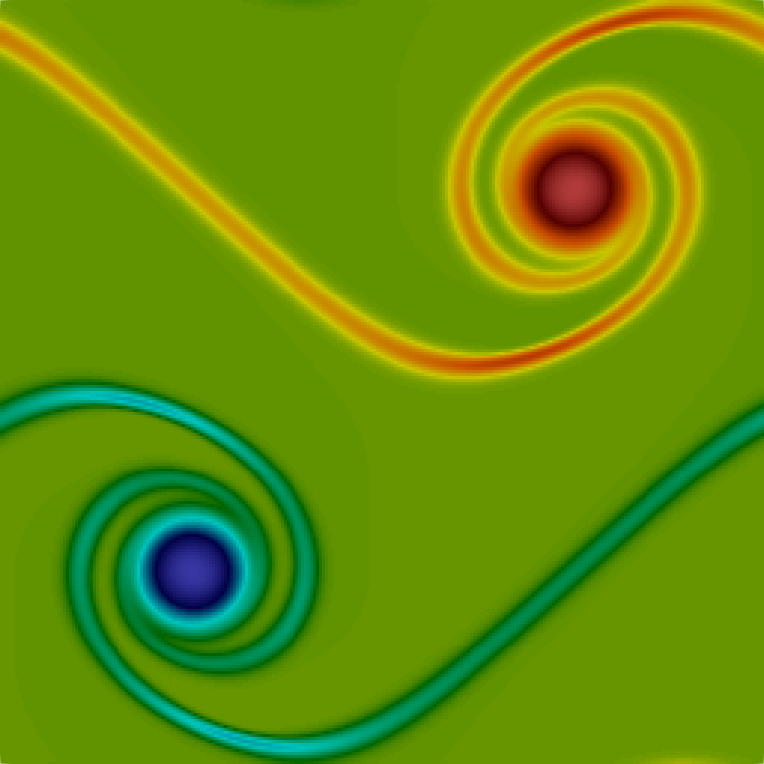}
	\begin{picture}(-4, 0)
		\put(-20,5){\footnotesize(j)}
	\end{picture}
	\includegraphics[scale=0.106]{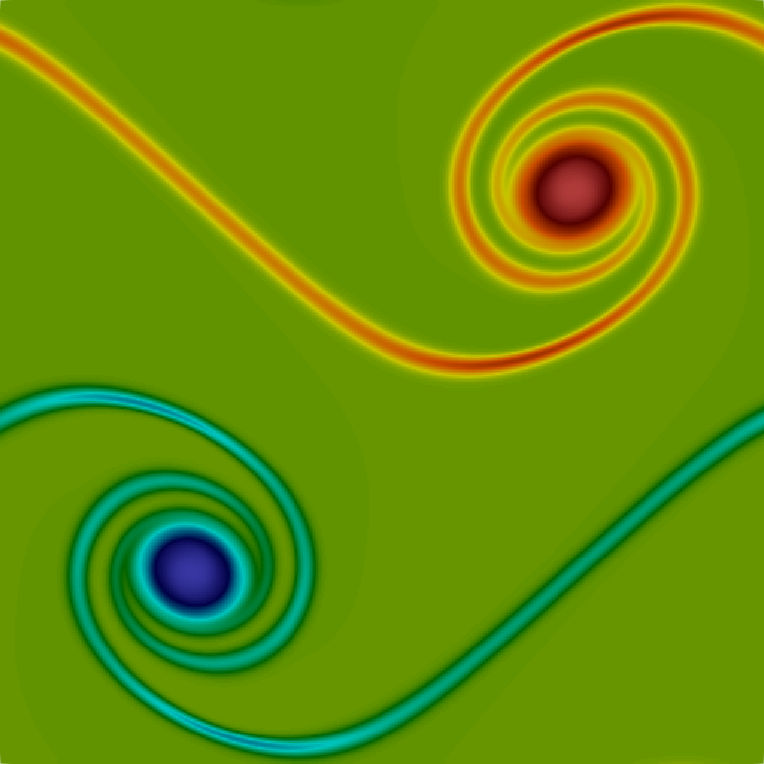}
	\begin{picture}(-4, 0)
		\put(-20,5){\footnotesize(o)}
	\end{picture}

	\caption{Normalized vorticity contours for the double periodic shear layer problem after unit characteristic time at an isothermal lattice temperature of $\theta=0.4$. Results are presented for the LBGK, regularized, uncorrected, partially-corrected and completely-corrected OReg schemes (top to bottom) on $N^2$ sized grids with $N$ = 128, 256 and 512 (left to right).\label{fig:dps-tp4-vorticity}}
\end{figure}
The parameters $\kappa=80$ and $\delta=0.05$ represent the width of the shear layers and the perturbation in the $y$-velocity that initiates a Kelvin-Helmholtz instability.
The system has a characteristic velocity $u_0=0.04$, and a Reynolds number Re = $(u_0 N)/\nu$ = 30000.
Simulations are run on three square grids of increasing size having side lengths $N$ = 128, 256 and 512 grid points.
Figures \ref{fig:dps-tp4-vorticity} and \ref{fig:dps-tp6-vorticity} represent vorticity contours at unit characteristic turnover time $t_c = N_\text{iter}/(N/u_0)$ for two isothermal temperatures $\theta$ = 0.4 and 0.6.

At $\theta = 0.4$, the uppermost panel of \Cref{fig:dps-tp4-vorticity} reveals that the LBGK scheme yields results that are unstable on the smallest grid and contain three distinct spurious vortices in each arm of the intermediate grid.
Upon further refinement of the grid, the observed numerical artifacts are seen to vanish.
We note that the LBGK scheme behaves in a similar manner even at the lattice reference temperature $\theta_0 = 1/3$, the only difference being the formation of a single spurious vortex in the arms of the shear layer on the intermediate grid.
This result is well known for the considered parameters and is therefore not presented here.
In contrast, it can be seen that the regularized, $\mathcal{O}(u/u^5)$ variable-viscosity uncorrected/partially corrected formulations and, the completely corrected OReg LB schemes yield stable results even on the coarsest grid.
Further, as with the LBGK scheme, the accuracy of the results obtained with each method improves as the grid resolution increases.
From \Cref{fig:dps-tp4-vorticity}, we also see that all the OReg schemes yield shear layers of progressively reducing thicknesses on grids having the same resolution.
This trend, which is particularly clear in the finest grid depicted by Figures \ref{fig:dps-tp4-vorticity}(m-o), clearly delineate the behaviour of the two proposed corrected OReg schemes and highlight the need for eliminating spurious contributions from the OReg pressure tensor at elevated isothermal temperatures.

However, it can be seen that the shear layers obtained with even the completely corrected OReg scheme are still thicker than those obtained using the LBGK and regularized LB schemes.
This behaviour manifests due to numerical errors arising from evaluating velocity gradients through \Cref{eq:geq-prefactor-evaluation}.
As discussed earlier, with accurate local representations of $\mathrm{X}_{\alpha\beta}$, the corresponding accuracy from  can be expected to improve.
Nevertheless, the utility of the proposed corrected OReg schemes can be appreciated in the $\theta=0.6$ case where both the LBGK and regularized LB schemes are found to be unstable even on grids as large as $N=2048$.
Note that such a grid resolution nearly corresponds to DNS resolution for Re = 30000.
As shown in \Cref{fig:dps-tp6-vorticity}, the OReg schemes are stable on all the three for grids considered with the behaviour of the uncorrected, partially and fully corrected OReg schemes being similar to those observed in the $\theta=0.4$ case.
\vspace{-0.5em}
\section{Conclusion}
\label{sec:conclusion}
\vspace{-1em}
\noindent
In this work, we demonstrate the utility of the OReg LB scheme in addressing the Navier-Stokes modelling errors arising on standard lattices in a fully local manner and present a framework that eliminates errors arising from violations of constraints on non-equilibrium populations.
The proposed framework deviates from existing strategies of extending the equilibrium distribution representation or introducing explicit source terms and modifies the non-equilibrium OReg populations with correction populations.
Additionally, by reducing the the number of non-equilibrium constraint violations being corrected, we demonstrate that partially corrected OReg schemes with significantly improved NS modelling accuracy can be also obtained.
We present realizations of partially and completely corrected OReg schemes and compare the proposed models against the Lattice BGK, regularized, and uncorrected OReg schemes in quasi one-dimensional and non-linear two-dimensional settings.
Among the OReg schemes, the completely corrected OReg scheme is, as expected,  found to yield the best accuracy followed by the partially corrected and the uncorrected OReg schemes.
We highlight that, while the present study is conducted in the specific context of the D2Q9 guided equilibrium representation in the absence of any external forces, the ideas presented in this work can be easily generalized to incorporate forcing schemes, 3D standard stencils and other equilibrium representations for advanced use cases such as, e.g., phase-field \citep{succiPhaseField2005} or fluctuating \citep{regFluclbm} LB models.
\vspace{-1em}

\begin{acknowledgments}
\vspace{-1em}
\noindent
Funded by the European Union - NextGenerationEU: This work has received funding from the European High Performance Joint Undertaking (J.U.) (Grant Agreement No. 101093169) and the Ministero delle Impresse e del Made in Italy (Grant No. P/040003/01-02/X64).
\end{acknowledgments}

\bibliography{generalizedOReg2D}

\end{document}